\documentclass[twocolumn,letterpaper]{IEEEAerospaceCLS} 

\usepackage{balance}  
\usepackage{cite}
\usepackage{graphicx}
\usepackage{float}
\usepackage{amsmath} 
\usepackage[utf8]{inputenc}
\usepackage[final]{pdfpages} 
\usepackage{pdfpages}
\usepackage{lipsum}
\usepackage{textcase}
\usepackage{url}
\usepackage{amsmath,esint}
\usepackage{epstopdf}
\usepackage{array} 
\usepackage{hyperref}
\usepackage{multirow}
\usepackage{subcaption}
\usepackage{booktabs}
\usepackage{enumitem}

\usepackage{url}

\newcolumntype{P}[1]{>{\centering\arraybackslash}p{#1}}
\newcolumntype{M}[1]{>{\centering\arraybackslash}m{#1}}

\pdfoutput=1
\begin{document}   
	
\title{UWB Air-to-Ground Propagation Channel Measurements and Modeling using UAVs}





\author{%
Wahab Khawaja\\ 
Department of ECE\\
North Carolina State University\\
Raleigh, NC 27606\\
wkhawaj@ncsu.edu
\and 
Ozgur Ozdemir\\ 
Department of ECE\\
North Carolina State University\\
Raleigh, NC 27606\\
oozdemi@ncsu.edu
\and 
Fatih Erden\\ 
Department of ECE\\
North Carolina State University\\
Raleigh, NC 27606\\
ferden@ncsu.edu 
\and 
Ismail Guvenc\\ 
Department of ECE\\
North Carolina State University\\
Raleigh, NC 27606\\
iguvenc@ncsu.edu 
\and 
David Matolak\\ 
Department of EE\\
University of South Carolina\\
Columbia, SC 29208\\
matolak@cec.sc.edu}

\maketitle

\begin{abstract} 
This paper presents an experimental study of the air-to-ground~(AG) propagation channel through ultra-wideband~(UWB) measurements in an open area using unmanned-aerial-vehicles (UAVs). Measurements were performed using UWB radios operating at a frequency range of 3.1~GHz~-~4.8~GHz and UWB planar elliptical dipole antennas having an omni-directional pattern in the azimuth plane and typical donut shaped pattern in the elevation plane. Three scenarios were considered for the channel measurements: (i) two receivers~(RXs) at different heights above the ground  and placed close to each other in line-of-sight~(LOS) with the transmitter~(TX) on the UAV and the UAV is hovering; (ii) RXs are in obstructed line-of-sight (OLOS) with the UAV TX due to foliage, and the UAV is hovering; and, (iii) UAV is moving in a circular path. Different horizontal and vertical distances between the RXs and TX were used in the measurements. In addition, two different antenna orientations were used on the UAV antennas (vertical and horizontal) to analyze the effects of antenna radiation patterns on the UWB AG propagation. From the empirical results, it was observed that the received power depends mainly on the antenna radiation pattern in the elevation plane when the antennas are oriented in the same direction, as expected for these omni-azimuth antennas. Moreover, the overall antenna gain at the TX and RX can be approximated using trigonometric functions of the elevation angle. The antenna orientation mismatch increases path loss, and produces a larger number of small powered multipath components~(MPCs) then when aligned. Similarly, additional path loss and a larger number of MPCs were observed for the OLOS scenario. In the case of the UAV moving in a circular path, the antenna orientation mismatch has smaller effects on the path loss than when the UAV is hovering, because a larger number of cross polarized components are received during motion. A statistical channel model for UWB AG propagation is built from the empirical results.


\end{abstract}


\tableofcontents

\section{Introduction}
The use of civilian unmanned aerial vehicles~(UAVs) for applications such as video recording, surveillance, search and rescue, and hot spot communications has seen a surge in recent years. The advantage of high mobility in air, ease in take-off/landing and operability, multiple flight controls, small size, and affordable prices compared to other aerial platforms make UAVs perfect candidates for numerous current and future applications. One of the promising applications is in the field of wireless communications, e.g., providing on-demand access to hot spot or disaster hit areas~\cite{wahab_survey}. A recent example was seen in Puerto Rico, after Hurricane Maria, where a majority of the cellular infrastructure was damaged. UAVs are used there by AT\&T as base stations to provide cellular coverage~\cite{COW}.   

\begin{figure}[!t]
	\centering
	\vspace{-.2cm}
	\includegraphics[width=\columnwidth]{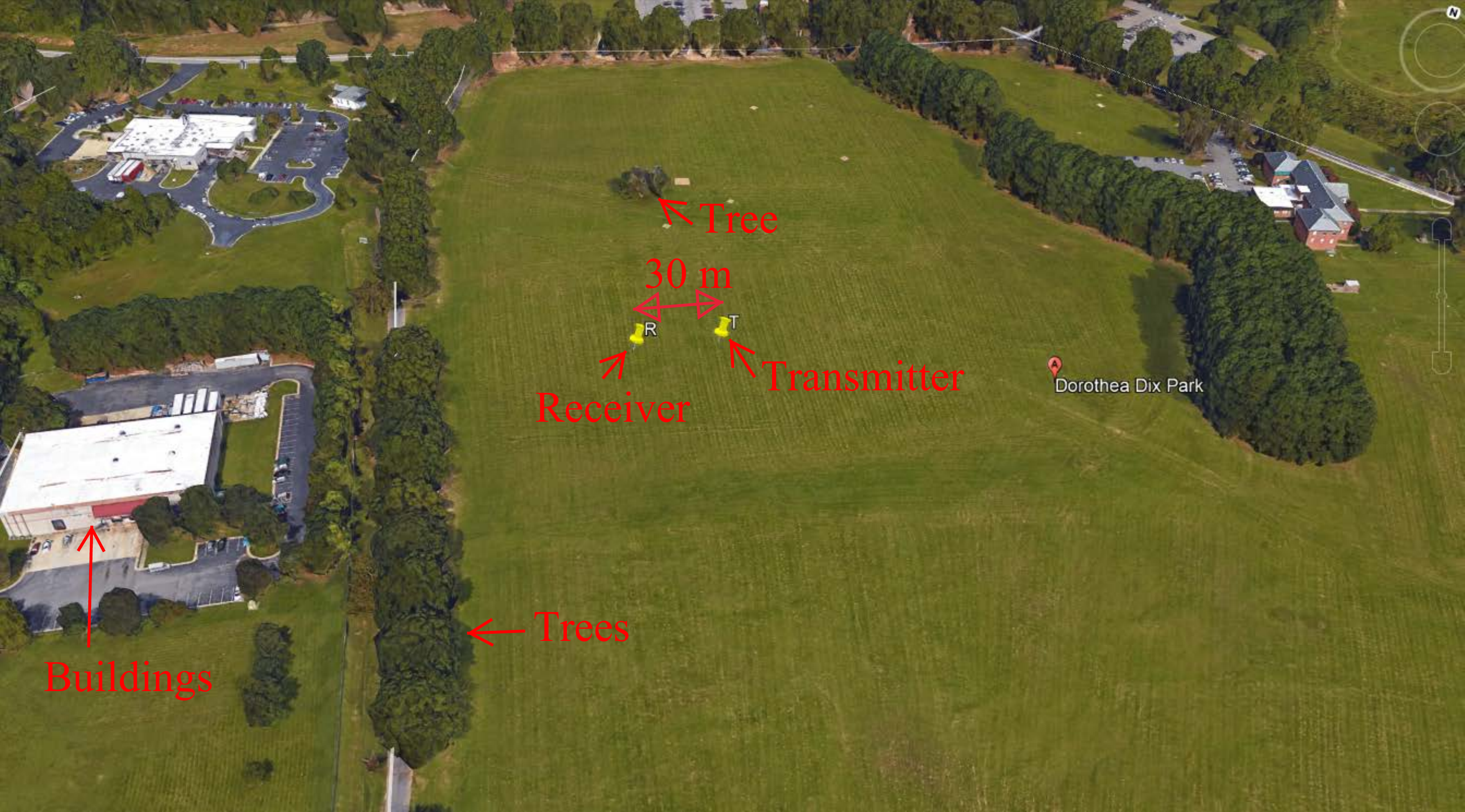}
	\vspace{-0.3cm}
	\caption{Channel measurement area from Google Earth.}\label{Fig:Google_maps}
\end{figure}

There are limited studies available on the air-to-ground~(AG) propagation channel characterization in the literature~\cite{wahab_survey}. Narrowband AG propagation channel measurements using UAVs in an urban environment~\cite{Narrow_band1} consider a Loo model~(Rice and lognormal) for signal variations. A two-ray path loss model in an urban environment is observed in~\cite{Narrow_band2} for narrowband measurements carried out in an urban environment. Wideband AG propagation channel measurement campaigns in the L-band and C-band were performed for over water, mountains and hilly area, suburban, and urban environments in~\cite{Matolak_over_water,Matolak_mountains_hilly,Matolak_Suburban_urban}. Large scale and small scale propagation channel statistics in the L-band and C-band were provided. Yet there are limited studies for AG propagation using UAVs that specifically focus on the antenna radiation effects~\cite{antenna_orien1, antenna_orien2, wahab_vtc}. 

To our best knowledge, there are also very limited ultra-wideband~(UWB) AG propagation channel measurements using UAVs in the literature, except for our previous studies~\cite{wahab_uwb,jianlin}. The large bandwidth of UWB radio signals allows high temporal resolution of multipath components~(MPCs) that can provide detailed information for a given propagation environment. Studying these MPC characteristics can help in understanding the AG propagation channel for future broadband communications~\cite{wahab_survey}.

In this study, we have carried out comprehensive channel measurements in an open area for three conditions: (1) line-of-sight~(LOS) path when the UAV is hovering without obstruction; (2) obstructed line-of-sight~(OLOS) path due to foliage within the link while the UAV is hovering; and, (3) UAV moving in a circular trajectory with unobstructed LOS path. A snapshot of the measurement area from Google Maps is shown in Fig.~\ref{Fig:Google_maps}. The measurements were conducted at different horizontal and vertical distances of the transmitter~(TX) on the UAV from the receivers~(RXs) on the ground. Two different antenna orientations, vertical and horizontal, were used at the TX, whereas the orientation of the antennas on the RXs were always vertical. The channel measurements were obtained using Time Domain P440 UWB radios operating in the frequency range $3.1$~GHz~-~$4.8$~GHz. All antennas are omni-directional. 

The main contributions of this AG measurement study can be summarized as follows: \\
1- The received power is mainly dependent on the antenna gain of the LOS component in the elevation plane when the antennas have the same orientation. Moreover, the antenna gain for the LOS component can be approximated by a trigonometric function of the elevation angle between the TX and the RX. \\
2- Antenna orientation mismatch results in higher path loss and larger number of small powered MPCs than when aligned.\\
3- The OLOS scenario due to foliage between the TX and the RX with the UAV hovering introduces additional attenuation and additional MPCs due to foliage.\\
4- The motion of the UAV in the circular path provides mitigation against antenna orientation mismatch effects in comparison to the aligned antenna orientation case. This mitigation is significant when compared to the UAV hovering without foliage scenario.\\
5- A larger number of MPC clusters are observed in the power delay profile~(PDP) for the UAV hovering without foliage scenario compared to the circularly moving UAV  scenario. \\
6- A statistical channel model is developed based on the empirical results.\\

The outline of this paper is as follows. Section~\ref{Section:Ch_analysis_model} covers the system model, UWB channel impulse response~(CIR) and antenna radiation pattern modeling. Section~\ref{Section:Ch_measurements} provides the channel measurement setup and description of the propagation scenarios. Empirical results are provided in Section~\ref{Section:Emp_results} and finally, Section~\ref{Section:Conclusions} concludes the paper.

 \section{System Model}\label{Section:Ch_analysis_model}

\subsection{Channel Impulse Response} \label{Section:CIR}
The CIR is modeled as a modified Saleh Valenzuela~(SV) model~\cite{saleh} given as
\begin{equation}
H(n)=\sum_{l = 0}^{N_{\rm C}-1}\sum_{\rm m=0}^{M_{l}-1}\alpha_{l,m} \exp\Big(j\varphi_{l,m}\Big)\delta\Big(n-T_l-\tau_{l,m}\Big), \label{Eq:Eq_CIR}
\end{equation}
where $N_{\rm C}$ is the total number of clusters, $M_{l}$ is the total number of MPCs within the $l^{\rm th}$ cluster, $\alpha_{l,m}$, and $\varphi_{l,m}$ are the amplitude and phase, respectively, of the $m^{\rm th}$ MPC of the $l^{\rm th}$ cluster. The mean square gain value of the $m^{th}$ MPC of the $l^{th}$ cluster is given in terms of the first as
\begin{equation}
\overline{\alpha_{l,m}^{2}} = \overline{\alpha_{0,0}^2}\exp({-T_l}{\eta})\exp({-\tau_{l,m}}{\gamma}), \label{Eq:Eq_Pwr} \end{equation}
 where $\overline{\alpha_{0,0}^2}$ is mean power gain of the first path of the first cluster, $\eta$ and $\gamma$ are the cluster and MPC power decay constants, respectively, $T_l$ and $\tau_{l,m}$ are the arrival times of the $l^{\rm th}$ cluster and $m^{\rm th}$ MPC of the $l^{\rm th}$ cluster, respectively.

The arrival of the clusters and MPCs within each cluster can be modeled by Poisson processes, with respective arrival rates, $\chi$ and $\varsigma$ observed during the excess delay window. The inter-arrival times of clusters and MPCs are independent and can be fitted with an exponential distribution function as:
\begin{align}
p(T_l|T_{l-1}) =& \chi \exp \big[-\chi(T_l - T_{l-1})\big], \label{Eq:Eq_Arrival_cluster}\\
p(\tau_m|\tau_{m-1}) =& \varsigma \exp \big[-\varsigma(\tau_m - \tau_{m-1}) \big].\label{Eq:Eq_Arrival_MPC}
\end{align}

\begin{figure}[!h]
	\centering
	\includegraphics[width=\columnwidth]{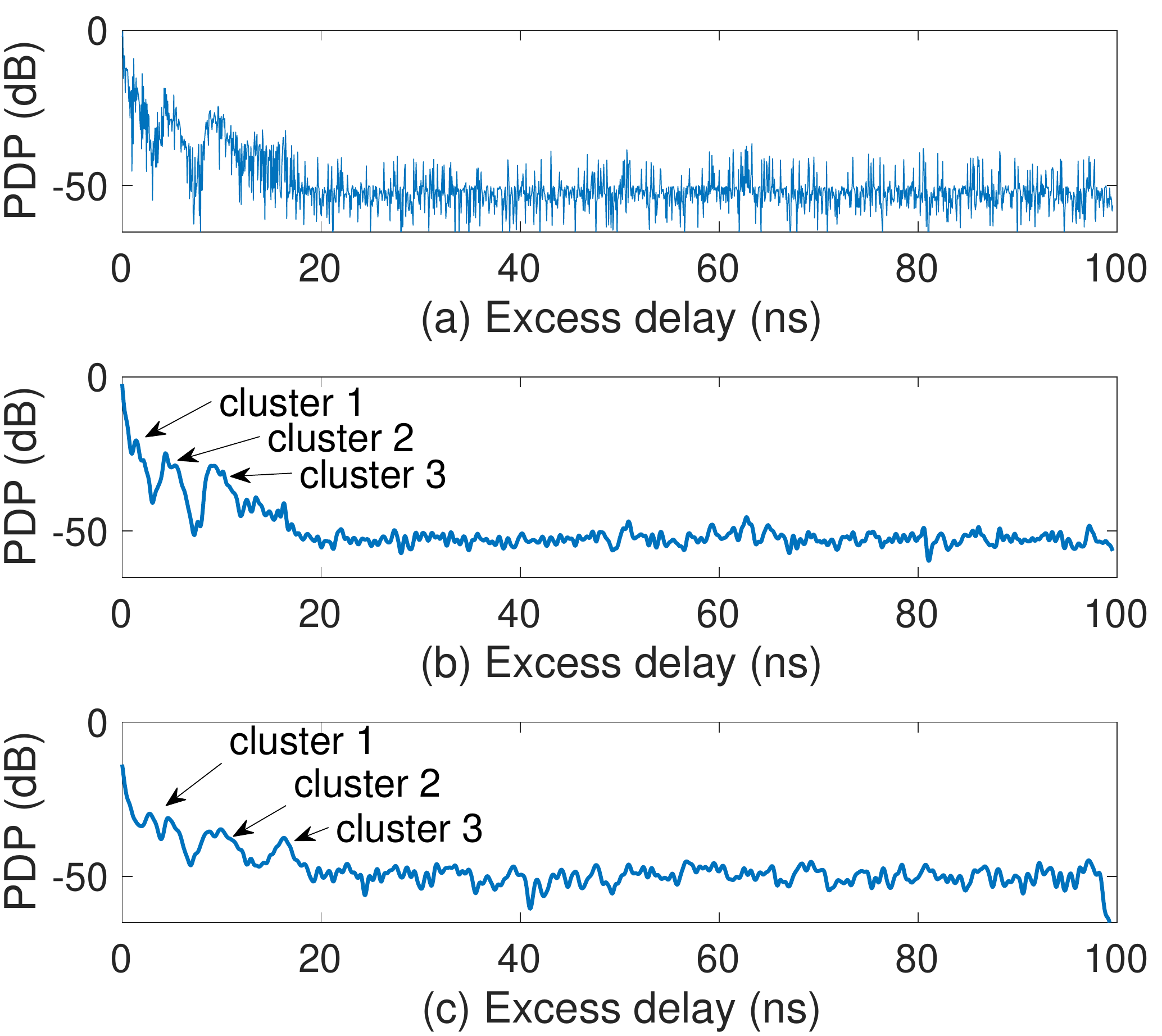}
	\caption{Normalized and averaged empirical PDP when the UAV is hovering at a height of $10$~m and horizontal distance $x = 30$~m from the receiver RX1 with (a) VV antenna orientation, (b) VV antenna orientation and smoothed version, (c) VH antenna orientation and smoothed version.}\label{Fig:PDP}
\end{figure}

The phase angle $\varphi$ is considered to be independent and uniformly distributed in the interval~$[0,2\pi)$. The PDP obtained from the CIR is used to analyze the MPC clusters shown in Fig.~\ref{Fig:PDP}. On the other hand, when the UAV is moving, the independence condition only partially holds, meaning that MPCs would have correlation for specific time lags. However, due to the large bandwidth (resulting in lower probability of superposition of MPCs from a given scatterer during mobility), the small number of surrounding scatterers and the moderate velocity of the UAV, the correlation among the MPCs will be weak.


\subsection{Antenna Radiation Pattern Modeling} \label{Section:Antenna_model}
\begin{figure}[!t]
	\centering
	\includegraphics[width=\columnwidth]{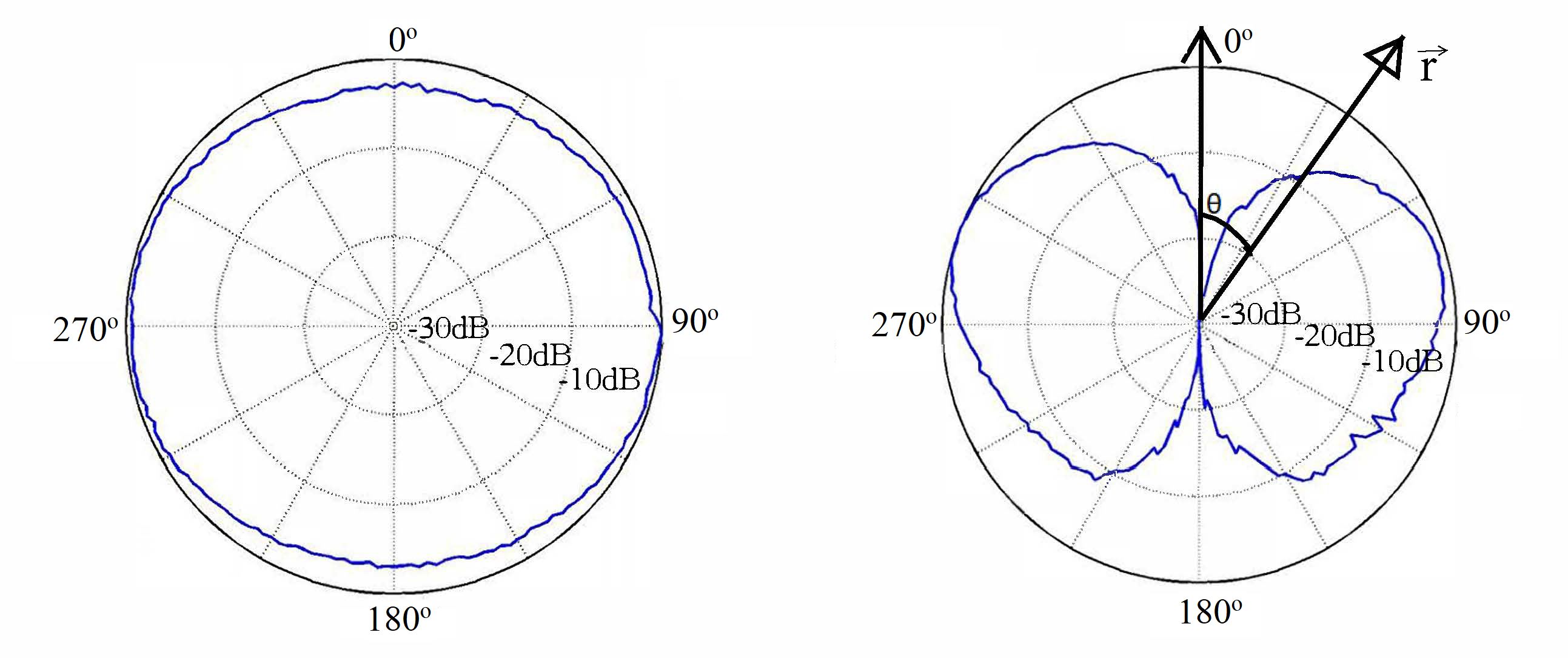}
	\caption{Antenna radiation pattern at $4$~GHz in the azimuth and elevation planes with directional vector at a given elevation angle.}\label{Fig:Antenna_pattern}
\end{figure}

\begin{figure}[!t]
	\begin{subfigure}{0.5\textwidth}
	\centering
	\includegraphics[width=\textwidth]{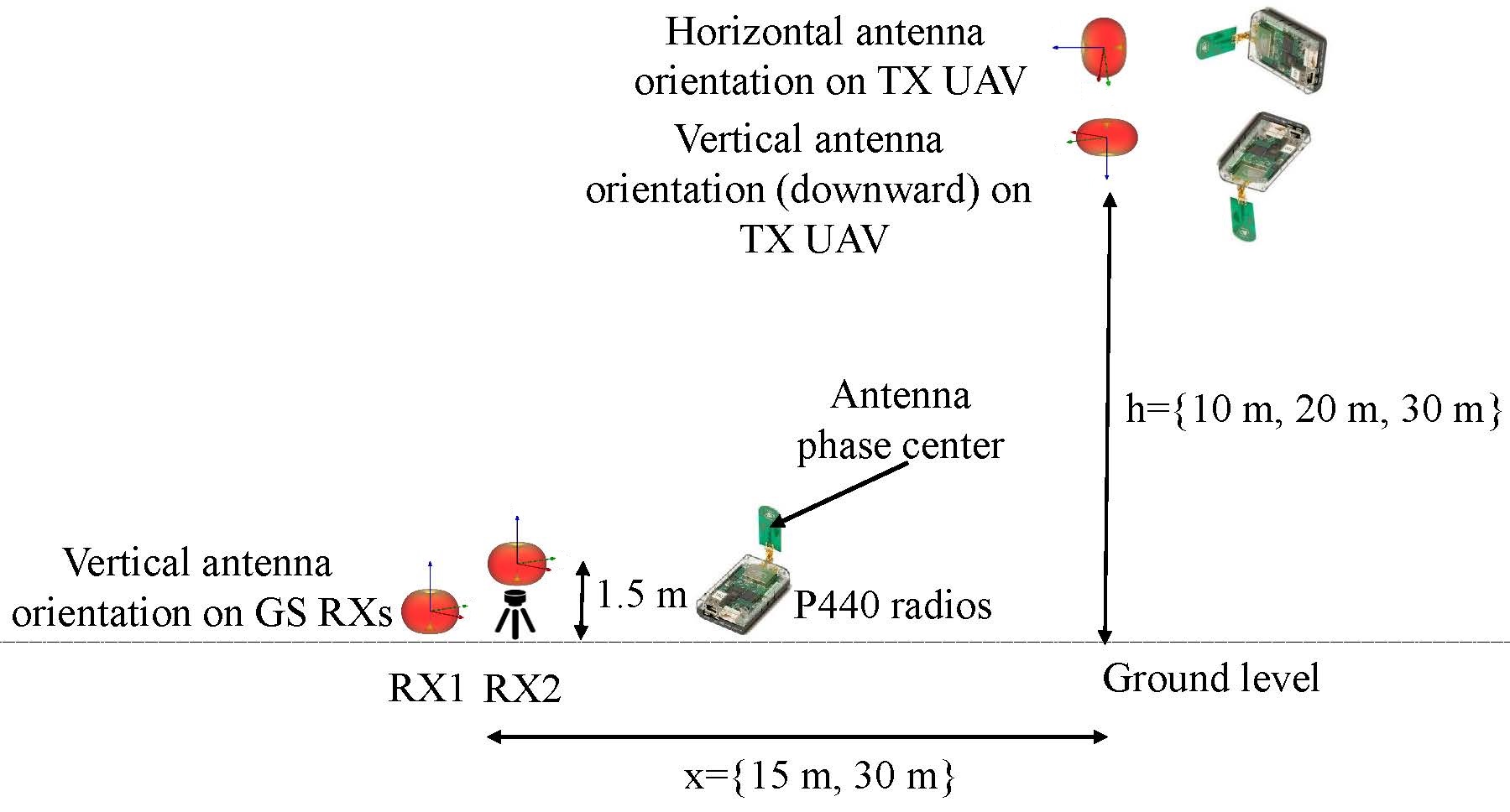} 
	\caption{}
    \end{subfigure}			
	\begin{subfigure}{0.5\textwidth}
	\centering
    \includegraphics[width=\textwidth]{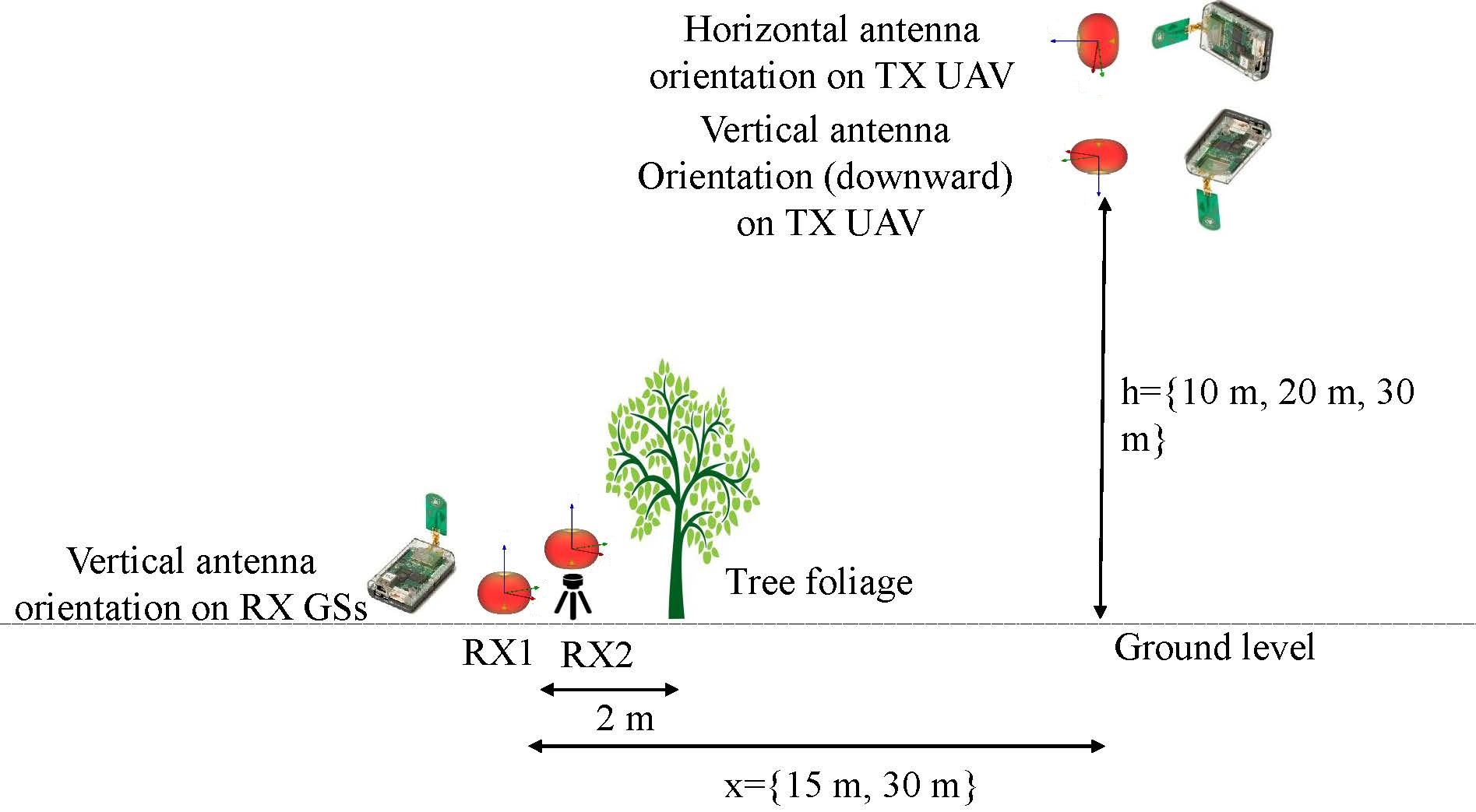}
	 \caption{}
     \end{subfigure}
     \begin{subfigure}{0.5\textwidth}
	\centering
     \includegraphics[width=\textwidth]{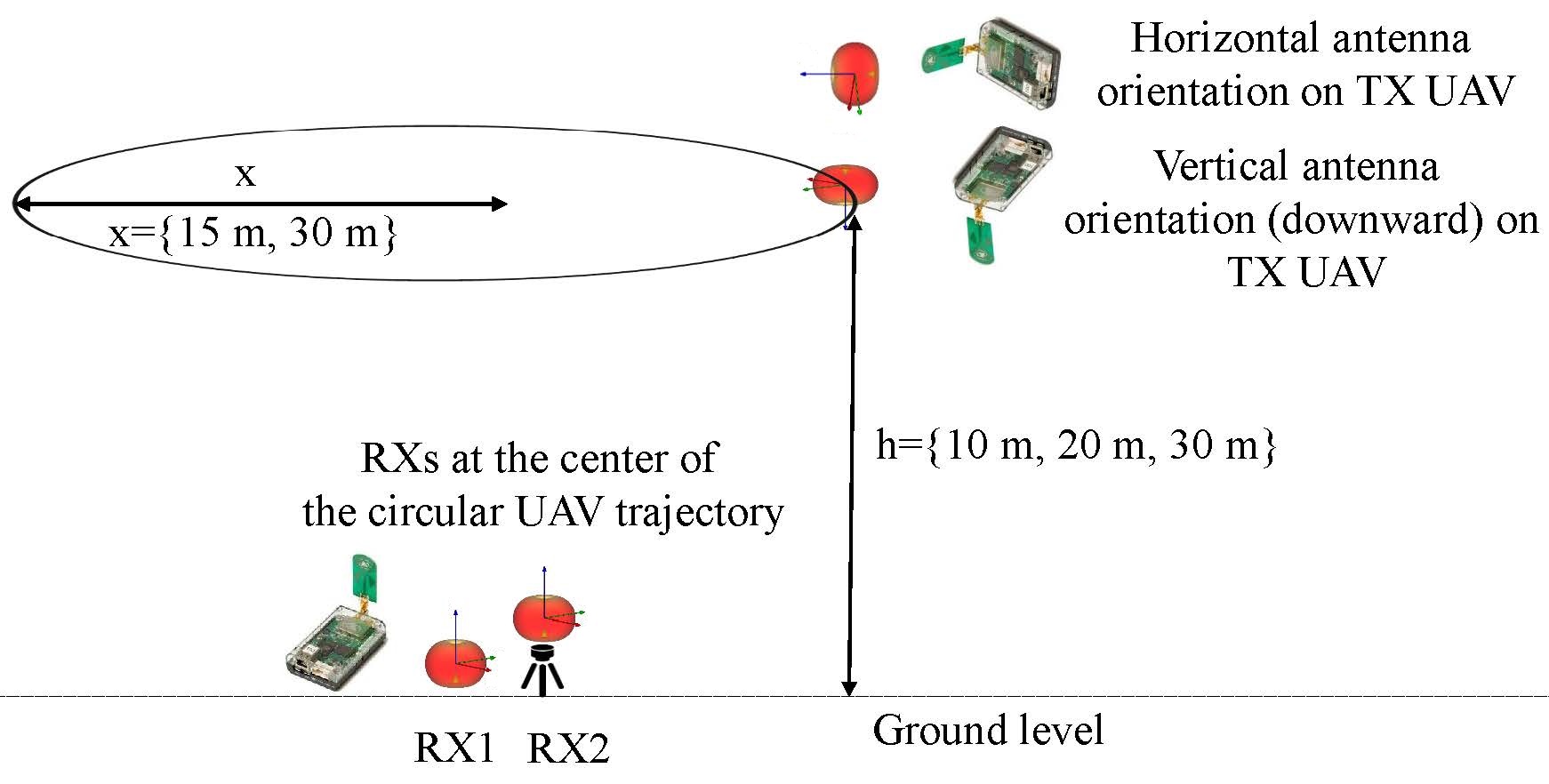}
	 \caption{}
     \end{subfigure}
     \caption{AG propagation scenarios in an open area for (a) UAV hovering without foliage, (b) UAV hovering with tree foliage obstruction, (c) UAV moving in a circular trajectory. }\label{Fig:Scenario_AG}
\end{figure}

BroadSpec UWB antennas from Time Domain are used in the experiment at both the TX and the RX. These antennas are planar elliptical dipoles with omni-directional pattern in the azimuth plane and a typical doughnut pattern in the vertical plane, shown in Fig.~\ref{Fig:Antenna_pattern}. The parameters of the antennas are provided in Table~\ref{Table:Sounding_parameters}. The vector $\vec{r}$ represents the direction of the link in the elevation plane at a given elevation angle~$\theta$ given by, $\theta=90-\tan^{-1}\Big(\frac{h}{x}\Big)$, where $x$ represents the horizontal distance between the RX and the TX, and $h$ represents the height of the UAV. Two antenna orientations are considered for the TX on the UAV~(downward and sideways). The antennas at the RXs are always pointing in the upward direction shown in Fig.~\ref{Fig:Scenario_AG}. For the first antenna orientation, the antenna at the TX is pointing vertically downward such that the antenna boresight~(with phase center in the middle) is facing the boresight of the RX antenna, when at the same height. This antenna orientation pair at the RX and TX is called vertical-vertical~(VV). For the second antenna orientation, the TX antenna is rotated $90^{\circ}$ sideways. We call this antenna orientation pair as vertical-horizontal~(VH). 

\begin{table}[!t]
	\begin{center}
		\caption{Specifications for channel measurements.}\label{Table:Sounding_parameters}
        \begin{tabular}{@{} |P{4cm}|P{3cm}| @{}}
			\hline
			\textbf{Parameter}&\textbf{Parameter value}\\			
			\hline
			Operating frequency band& $3.1$~GHz~-~$4.8$~GHz \\
            \hline
            Center frequency& $4.3$~GHz \\
            \hline
            Pulse duration& $1$~ns \\
            \hline
            Transmitted power& $-14.5$~dBm \\
            \hline
            Dynamic range& $48$~dB \\
            \hline
            Pulse repetition rate& $10$~MHz \\
            \hline
            Noise figure at RX& $4.8$~dB \\
            \hline
            RX sensitivity& $-104$~dBm \\
            \hline
            RX time bin resolution& $1.9073$~ps \\
            \hline
            RX waveform measurements at& $32$ time bin interval \\
            \hline
            Communication link&Packet communication \\
            \hline 
            Antenna type at TX and RX& Planar elliptical dipole \\
            \hline
            Antenna gain& $3$~dBi \\
            \hline
            Polarization& Vertical \\
            \hline
            Antenna pattern& Omni-directional in azimuth plane~($\pm 1.5$~dB) \\
            \hline
            Voltage standing wave ratio& $1.75:1$ \\
            \hline
            Antenna phase response& Linear \\
            \hline
            
	\end{tabular}
		\end{center}
\end{table}

\begin{figure}[!t]
	\centering
	\includegraphics[width=\columnwidth]{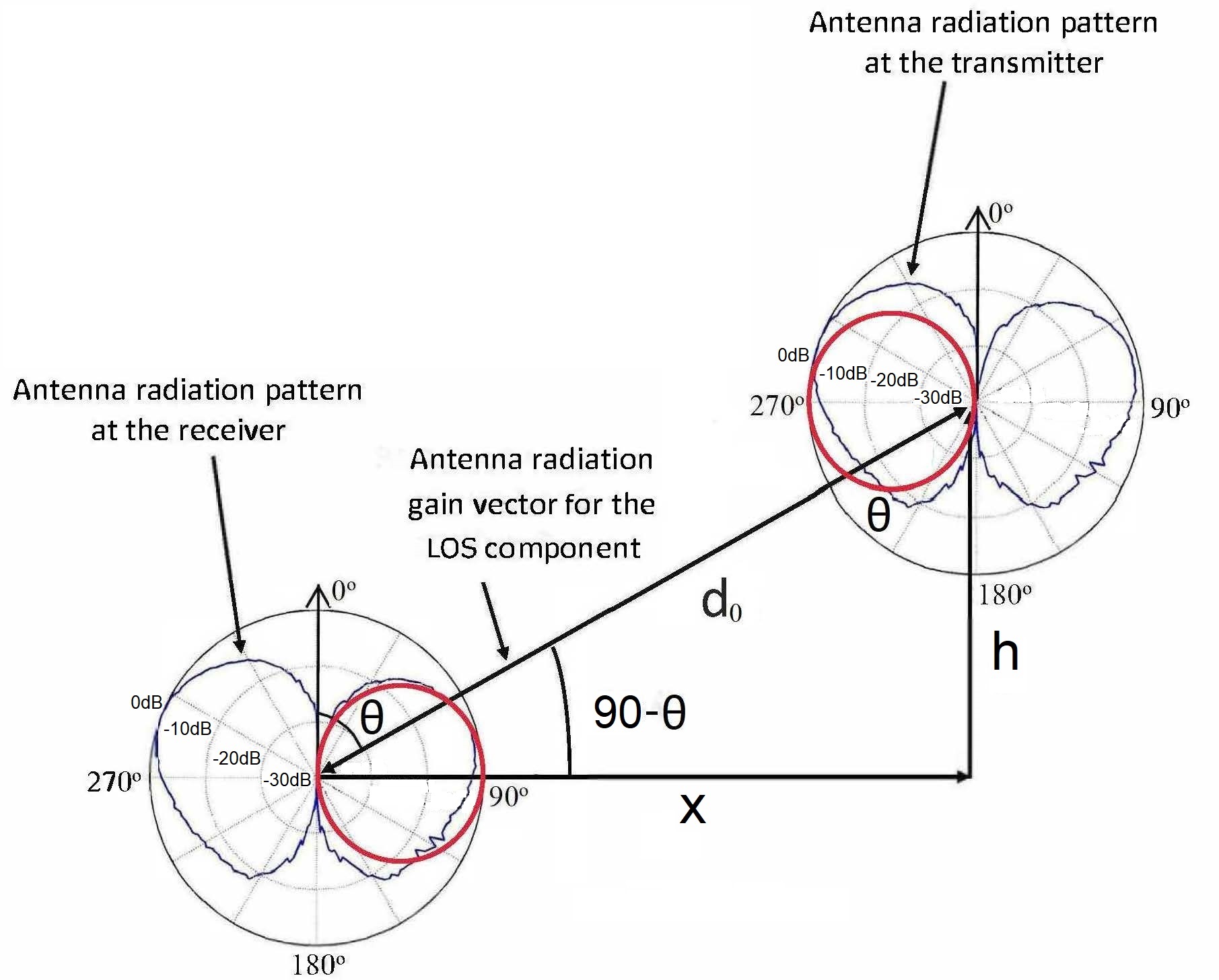}
	\caption{Antenna radiation pattern in the elevation plane at $4$~GHz for the transmitter on the UAV and receiver on the ground station.}\label{Fig:TX_RX_Ant_pattern}
\end{figure}

In AG propagation, it is important to consider the antenna radiation pattern in the three dimensions~\cite{wahab_survey}. The antenna radiation pattern in the elevation plane plays a key role in determining the received power especially, at higher elevation angles. The elevation angle is defined as per a line directly connecting TX to RX (along the LOS component, if present). The link in Fig.~\ref{Fig:TX_RX_Ant_pattern} represents the LOS component from the phase center of the TX antenna to the RX antenna. This LOS component makes an angle $\theta$ with the vertical axis. This vector has the highest value when angle $\theta$ is around $90^\circ$ or $270^\circ$, whereas it has minimum value near the vertical axis. 

\begin{figure}[!t]
	\centering
	\includegraphics[width=\columnwidth]{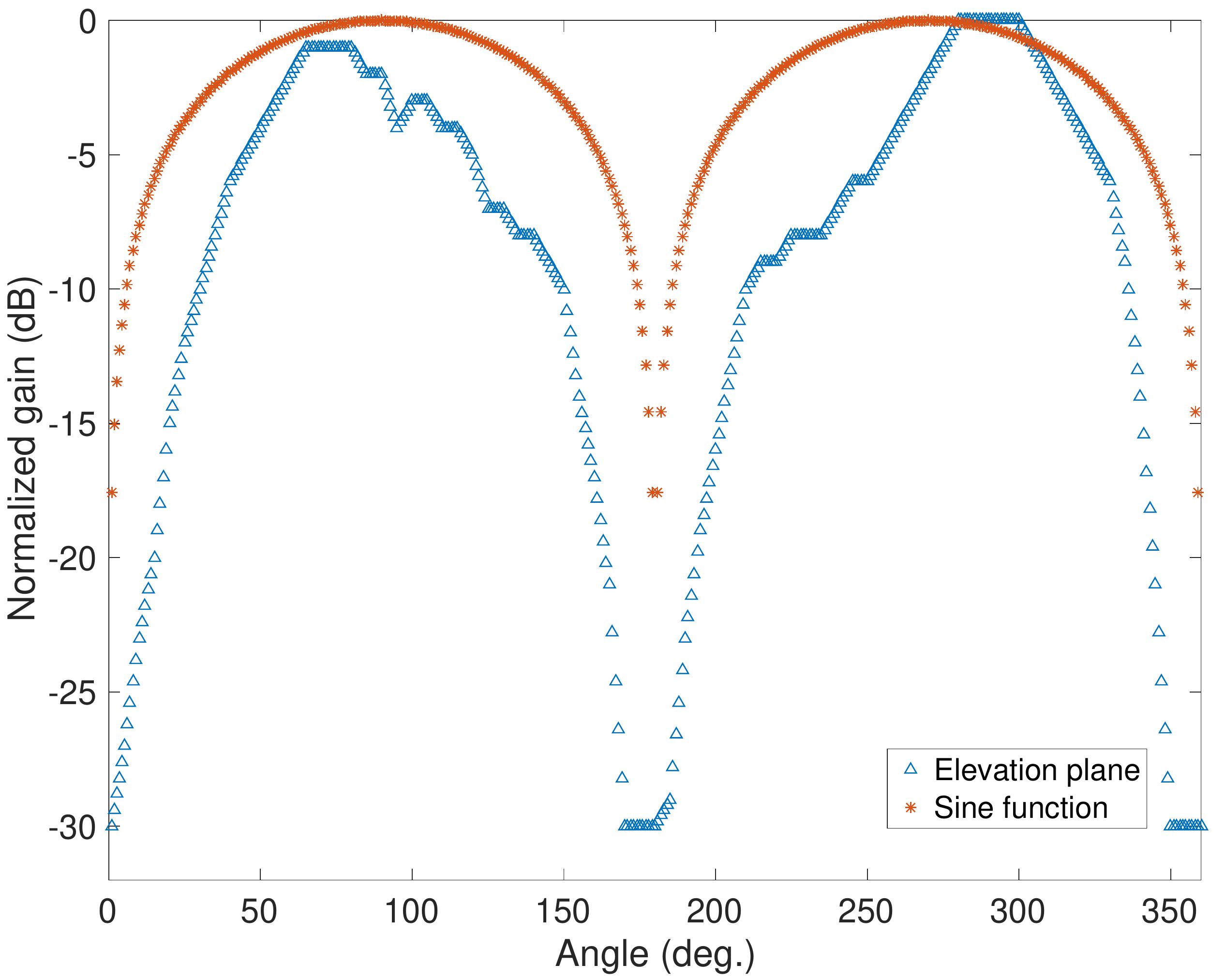}
	\caption{Comparison of normalized antenna gain at $4$~GHz in the elevation plane with the absolute sine function.}\label{Fig:Elevation_gain_sine}
\end{figure}

Moreover, the antenna radiation pattern is approximately symmetrical around the vertical axis as shown by the red circles in Fig.~\ref{Fig:TX_RX_Ant_pattern}. Therefore, the antenna gain for the LOS component in the elevation plane can be approximated by an absolute sine trigonometric function, i.e., $|\sin  \theta|$. Comparison of normalized antenna gain in the elevation plane with the absolute sine function $|\sin  \theta|$ over an angular span of [$0$~$360^\circ$] is provided in Fig.~\ref{Fig:Elevation_gain_sine}. For the VV antenna orientation pair, the overall antenna gain for the LOS component can be approximated as $\sqrt{|\sin\theta|~|\sin\theta|}$ . Hence as ${\theta}$ decreases, the LOS component is attenuated

\subsection{Antenna Polarization Mismatch Losses} \label{Section:Polarization} 
\begin{figure}[!t]
	\centering
	\includegraphics[width=\columnwidth]{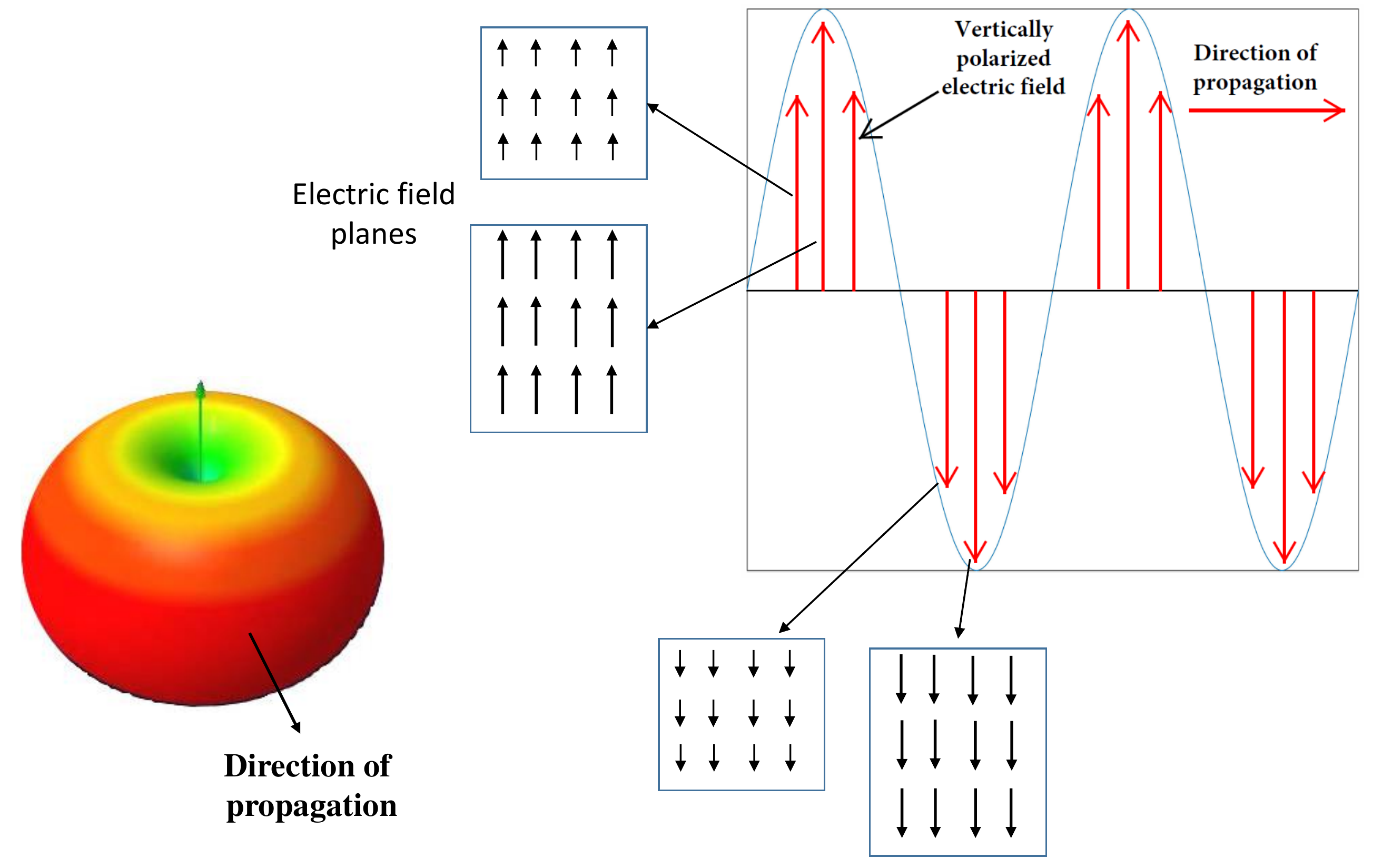}
	\caption{Vertically polarized electric field propagation.}\label{Fig:Electric_field_propag}
\end{figure}

 In order to observe the polarization mismatch phenomenon for omni-directional AG propagation, let us consider Fig.~\ref{Fig:Electric_field_propag}, here, each plane represents the respective electric field direction and strength at a given instance in the far field. If the electric field planes at the RX are aligned to those incident from the TX, there is no polarization mismatch loss. This means that the polarization vectors~(direction of electric field variation) at the TX and RX are in the same direction. However, for the VH antenna orientation pair, the incident and received electric field planes are not aligned, resulting in reduced received power due to polarization mismatch~\cite{Pol_loss}. In the ideal case, the VH antenna orientation pair should yield no reception for the linearly polarized antennas, as the electric field direction at the TX and RX are orthogonal to each other. However, due to non-ideal cross-polarization discrimination, and to reflections from scatterers in the environment, cross polarized components appear. These components help in the reception for the VH antenna orientation pair. 
 
 \subsection{Received Power} \label{Section:Rec_pwr}
 If $s(n)$ is the transmitted signal, then the received signal is given by $ R(n)=s(n)\odot H(n)$, where $\odot$ is the convolution operation. The received signal consists of LOS and non-line-of-sight~(NLOS) components given as:
 \begin{equation}
    \begin{split}
    \begin{aligned}
    R_m(n)&= {\rm Re}\bigg[\frac{\lambda \Gamma_m(\phi_m,\theta_m)}{4\pi d_m}\sqrt{G_{\rm T}(\phi_m^{\rm TX},\theta_m^{\rm TX}) G_{\rm R}(\phi_m^{\rm RX},\theta_m^{\rm RX})}\\ &~~~~s(n-\tau_m)\exp \bigg(\frac{-j2\pi d_m}{\lambda}\bigg) \bigg],
\end{aligned}
\end{split}
\end{equation}
where $m=0,1,2,.....M$, $m=0$ represents the LOS component, $\lambda$ is wavelength of the transmitted signal, $\Gamma_m(\phi_m,\theta_m)$ is the reflection coefficient of the $m^{\rm th}$ component, with $\phi_m$ and $\theta_m$ being the azimuth and elevation angles of the received components with the scatterers~(considering first order reflections), $G_{\rm T}(\phi_m^{\rm TX},\theta_m^{\rm TX})$ is the gain of the antenna at the TX  at respective azimuth and elevation angles of departures, $G_{\rm R}(\phi_m^{\rm RX},\theta_m^{\rm RX})$ is the gain of the antenna at the RX  at respective azimuth and elevation angles of arrivals, and finally $\tau_m$ and $d_m$ are the delay and distance of $m^{\rm th}$ component, respectively. We use the terms reflection and scattering mostly interchangeably here, understanding that these represent distinct physical propagation mechanisms; their aggregate effect is captured by $\Gamma$ in our formulation.

For the LOS component $\Gamma_0(\phi_0,\theta_0) = 1$, and the distance of the path between the TX and RX will be $d_0$, shown in Fig.~\ref{Fig:TX_RX_Ant_pattern}. Similarly, $\tau_0 = 0$, for the LOS component in our case. As noted, the gain of antenna for the LOS component can be approximated by the trigonometric function discussed in Section~\ref{Section:Antenna_model}. Therefore, for the VV antenna orientation pair, the LOS received component can be represented as

\begin{equation}
    \begin{split}
    \begin{aligned}
      R_0(n)&={\rm Re}\bigg[{ \frac{\lambda}{4\pi d_0}\sqrt{|\sin\theta||\sin\theta|}s(n)\exp \bigg(\frac{-j2\pi d_0}{\lambda}\bigg)} \bigg]. \label{Eq:Received_power}
    \end{aligned}
     \end{split}
\end{equation}

The received power is calculated from the LOS and NLOS components by time averaging, given as
 \begin{equation}
 \begin{aligned}
 P_{\rm R}&={\rm E}\bigg[\big|R_0(n)\big|^2\bigg] + {\rm E}\bigg[\sum_{m=1}^{M-1} \big|R_m(n)\big|^2\bigg]\\
 P_{\rm R}&= P_{\rm R}^{\rm LOS} + P_{\rm R}^{\rm NLOS}, \label{Eq:RX_pwr}
\end{aligned}
 \end{equation}
 where $P_{\rm R}^{\rm LOS}$ and $P_{\rm R}^{\rm NLOS}$ are the time average power of the LOS and the NLOS components, respectively. The received power $P_{\rm R}$ is scaled down by the polarization loss factor in the VH antenna orientation due to polarization misalignment losses. 

\section{Channel Measurement Setup}\label{Section:Ch_measurements}
In this section, we explain the channel measurement setup conducted using Time Domain P440 radios and DJI Phantom~4 UAV. The measurements are carried out in an open field in North Carolina. A Google map image of the measurement area is shown in Fig.~\ref{Fig:Google_maps}. 

\subsection{Channel Sounding with Time Domain P440 UWB Radios}
Channel sounding equipment is generally very bulky and often requires wired synchronization. This puts a constraint on the AG propagation channel measurements with conventional channel sounders using UAVs. Therefore, we used Time Domain P440 radios for UWB channel sounding since they provide easy to setup bi-static channel measurements. Additionally, no physical connection is required for synchronizing the TX and the RX. A central synchronizing clock signal is sent from the TX to the RX through packets. A very narrow pulse similar to a Gaussian shape in the time domain is used. The duration of each pulse is $1$~ns and the repetition interval of the pulse is $100$~MHz resulting in a scan duration of $100$~ns. The pulses are integrated into customize sized packets.

Due to coherent operation of TX and RX, the signal to noise ratio~(SNR) can be adjusted by changing the integration of pulses per packet. By increasing the pulse integration per packet, we can achieve longer ranges due to higher SNR. This can help in overcoming the power emission limitations by FCC\cite{FCC_emission}. However, higher pulse integration will lead to lower data rates, resulting in fewer number of channel scans captured in a given timing window. In our experiment, we have used a pulse integration of $1024$ pulses per packet. This value ensures that we capture channel scans in a timing window without significant change of the propagation channel and at a reasonable link distance. 

In addition to emission requirements by the FCC for UWB, there are two main factors affecting the SNR of the received signal. These factors limit the extraction of the CIR using the CLEAN algorithm, which requires a given threshold of SNR. First is the preference to use omni-directional antennas compared to directional antennas for AG communications, and the second is the measurement noise. Omni-directional antennas with very small antenna gain are expected to be affected more from the surrounding environment variations. These variations may be larger for aerial platforms than for terrestrial. Second, we observed high measurement noise by using the equipment on-board UAVs compared to that observed at the ground stations~(GSs). This is mainly due to noise generated from the propellers, vibrations on-board the UAV, and ambient effects, e.g., high temperatures experienced on-board the UAV, especially at higher UAV heights during a sunny day. These factors increase the RX noise, causing more frequent loss of transmitted packets, hence requiring larger coherent pulse aggregation per packet.

The UWB radios used in the experiment operate in the bi-static mode with a single transmit and receive antenna. In this mode, the TX continuously sends packets at an inter-packet delay of $10$~ms. A rake RX is used with a delay bin resolution of $1.9073$~ps. A standard $32$~bin duration is maintained between two measurements i.e., each measurement sample is processed after $61$~ps. The operating frequency range is $3.1$~GHz~-~$4.8$~GHz with an effective bandwidth of $1.7$~GHz~\cite{Time_Domain}.

\begin{figure}[!t]
	\centering
	\includegraphics[width=\columnwidth]{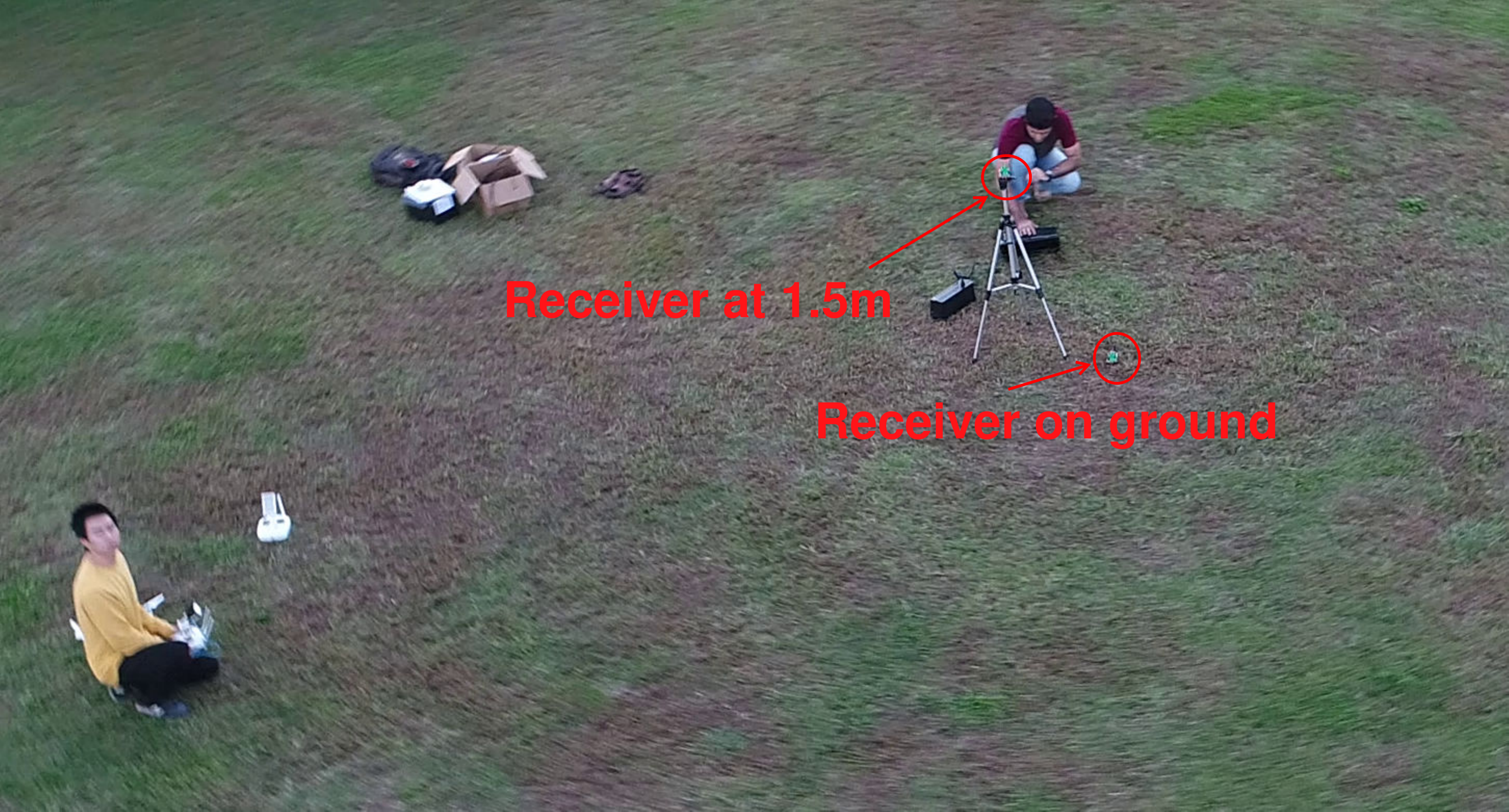}
	\caption{Channel measurements using DJI Phantom UAV and UWB P440 radios at two RX positions (snapshot from the UAV).}\label{Fig:Scenario_original}
\end{figure}

The UAV used for the measurements was a DJI Phantom~4. Using the GS auto-pilot application\cite{DJI_GS}, the UAV flew exactly at the designated flight coordinates. A snapshot of the measurement environment is shown in Fig.~\ref{Fig:Scenario_original}.

\begin{figure}[!t]
	\centering
	\includegraphics[width=\columnwidth]{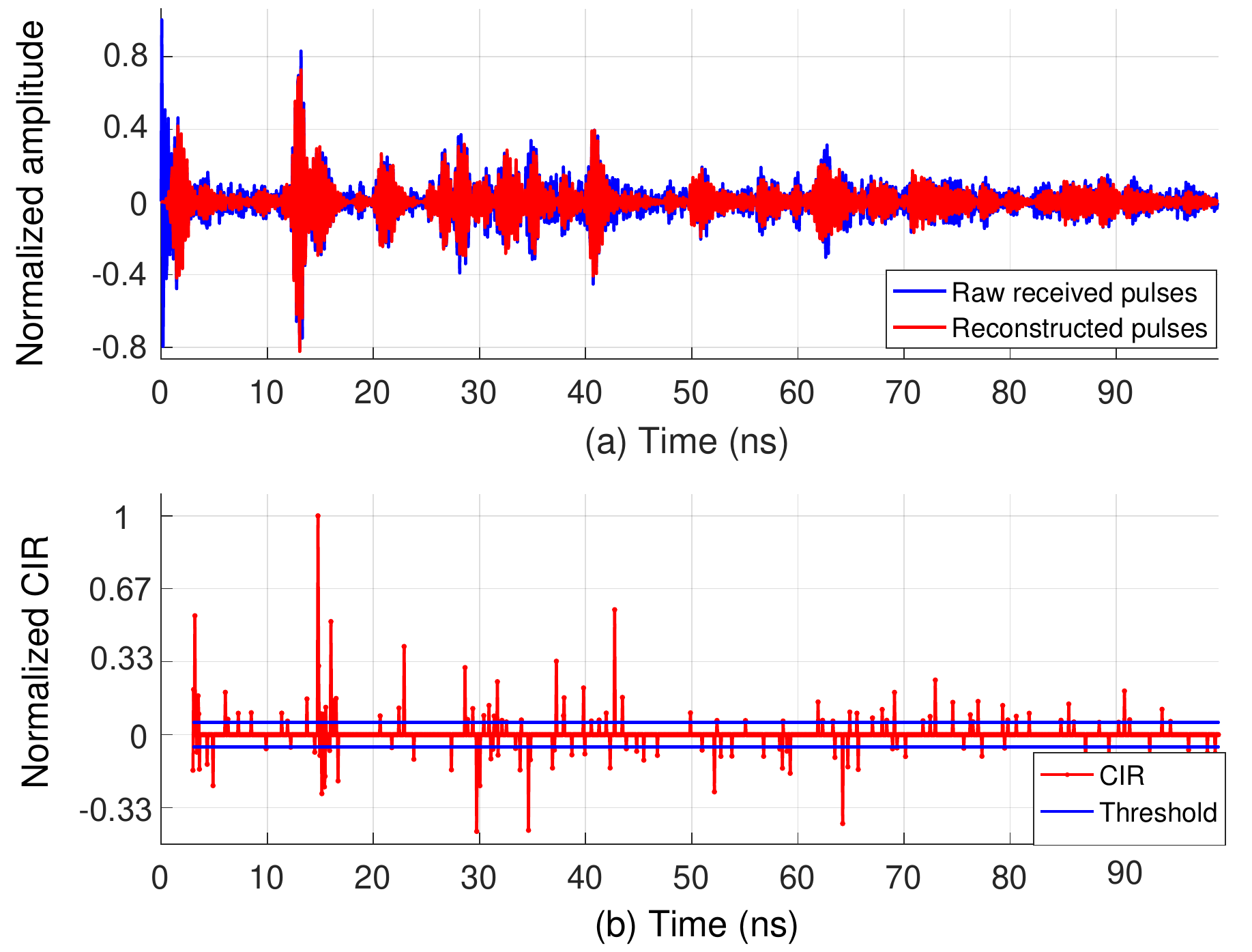}
	\caption{(a) Raw received and reconstructed pulses at the RX, and (b) channel impulse response with respect to time obtained from raw received pulses at the RX.}\label{Fig:pulse_tx_rx}
\end{figure}

The received raw pulses are shown in Fig.~\ref{Fig:pulse_tx_rx}(a) in blue, whereas the reconstructed pulses shown in red are obtained by convolving the CIR shown in Fig.~\ref{Fig:pulse_tx_rx}(b) with the template waveform. The CIR in Fig.~\ref{Fig:pulse_tx_rx}(b) is obtained by deconvolving the received pulses with the template waveform. The blue horizontal lines indicate the amplitude threshold of the MPCs selected at $20\%$ of the input signal. The channel sounding parameters are provided in Table~\ref{Table:Sounding_parameters}.


\subsection{Propagation Scenarios for Measurements}  
The experiments are designed to explore the UWB AG propagation channel characteristics in an open area. Three considered propagation scenarios are shown in Fig.~\ref{Fig:Scenario_AG}. For the first scenario shown in Fig.~\ref{Fig:Scenario_AG}(a), there is no obstruction between the TX and the RX direct path while the UAV is hovering. For the second OLOS scenario, the TX and RXs are placed such that there is a medium sized tree of height approximately $8$~m in between them shown in Fig.~\ref{Fig:Scenario_AG}(b). The branches and leaves of the tree partially obstruct, scatter and diffract the transmitted energy. In the third scenario, measurements are taken while the UAV is moving in a circle, with RXs at the center shown in Fig.~\ref{Fig:Scenario_AG}(c). The velocity of the UAV is set at $6.1$~m/s. The motion in a circle ensures that distance remains constant between the TX and the RXs.

In all three propagation scenarios, two antenna orientations pairs, VV and VH were used for the three UAV heights of 10~m, 20~m, and 30~m at two horizontal distances of $x=15$~m and $x=30$~m. Two RXs, RX1 and RX2 were placed close to each other at heights of $10$~cm and $1.5$~m, respectively, from the ground.

\section{Empirical Results} \label{Section:Emp_results}
In this section, empirical results are presented for different propagation scenarios. This includes results for path loss and small scale channel analysis of MPCs.
\subsection{Path Loss} \label{Section:PL}
\begin{figure*}[!t]
	\centering
	\includegraphics[width=\textwidth]{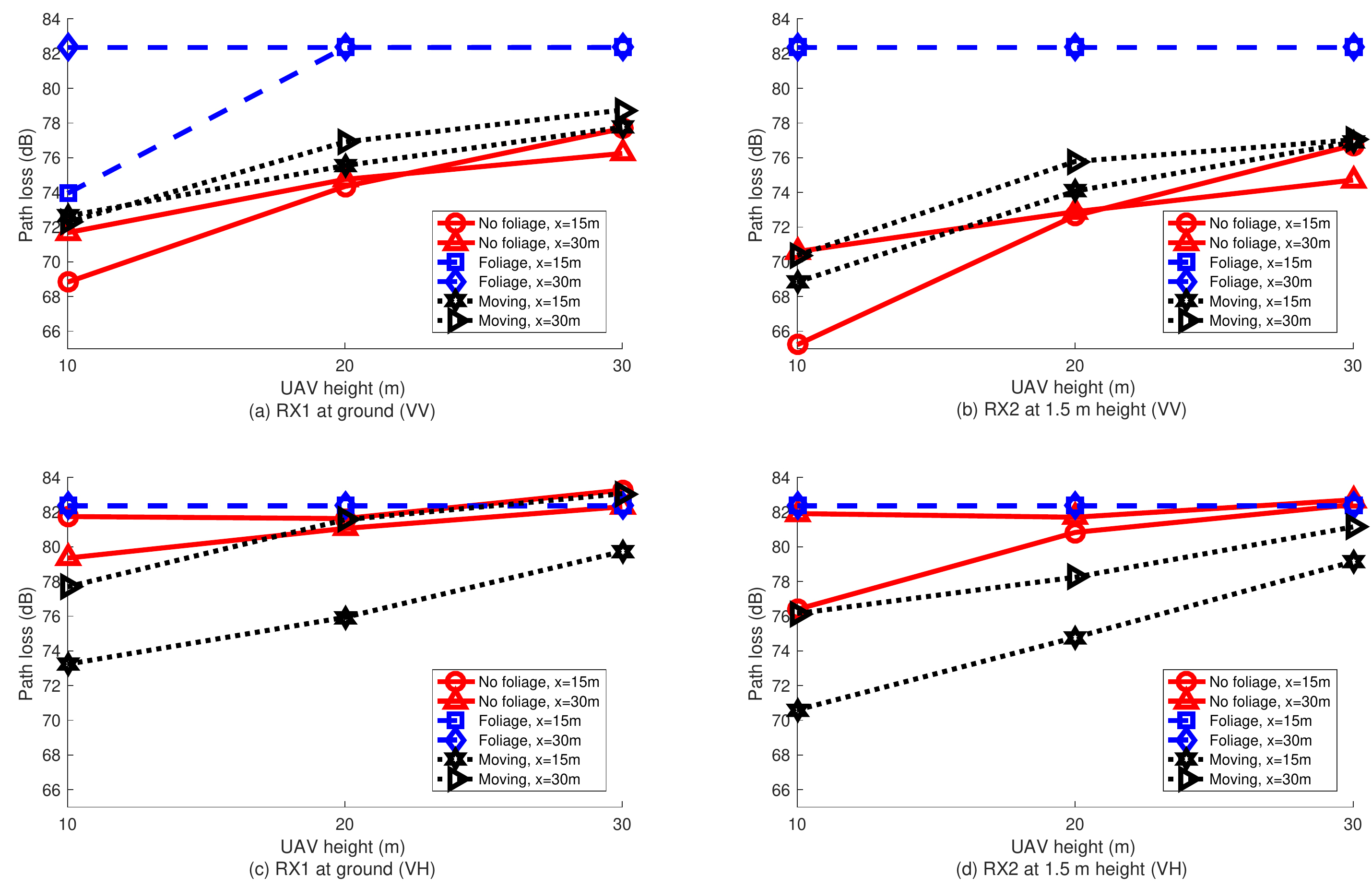}
	\caption{Path loss averaged over the channel scans for UAV hovering without foliage, with foliage obstructed, and UAV moving in a circle scenarios. Measurements are at two horizontal distances, $x=15$~m, $x=30$~m and three UAV heights of $10$~m, $20$~m, $30$~m, respectively. (a) RX1 at ground and VV antenna orientation pair, (b) RX2 at height of $1.5$~m and VV antenna orientation pair, (c) RX1 at ground and VH antenna orientation pair, (d) RX2 at height of $1.5$~m and VH antenna orientation pair.}\label{Fig:PL_mean}
\end{figure*}

The empirical path loss $L(d)$ in dB is evaluated as
\begin{equation}
L(d)~[\rm dB] = 10\log_{10}{L(d_{\rm ref})} + 10\log_{10} \frac{P^{d_{\rm ref}}}{P^{d}},
\end{equation}
where $L(d_{\rm ref})=(\frac{4\pi d_{\rm ref}}{\lambda})^2$, is the free space path loss at reference distance $d_{\rm ref} = 1$~m~\cite{rappaport}, and $\lambda$ corresponds to the wavelength at the center of the UWB signal spectrum, $P^{d_{\rm ref}}$ and $P^{d}$ are the received signal powers~(averaged over time) at reference distance $d_{\rm ref}$ and distance $d$, respectively. The distance $d$ corresponds to the respective horizontal distance of the the TX from the RX and height of the TX. This distance is also equal to the LOS component distance shown in Fig.~\ref{Fig:TX_RX_Ant_pattern}. Fig.~\ref{Fig:PL_mean} shows the path loss for different propagation scenarios.

\begin{figure}[!t]
	\centering
	\includegraphics[width=\columnwidth]{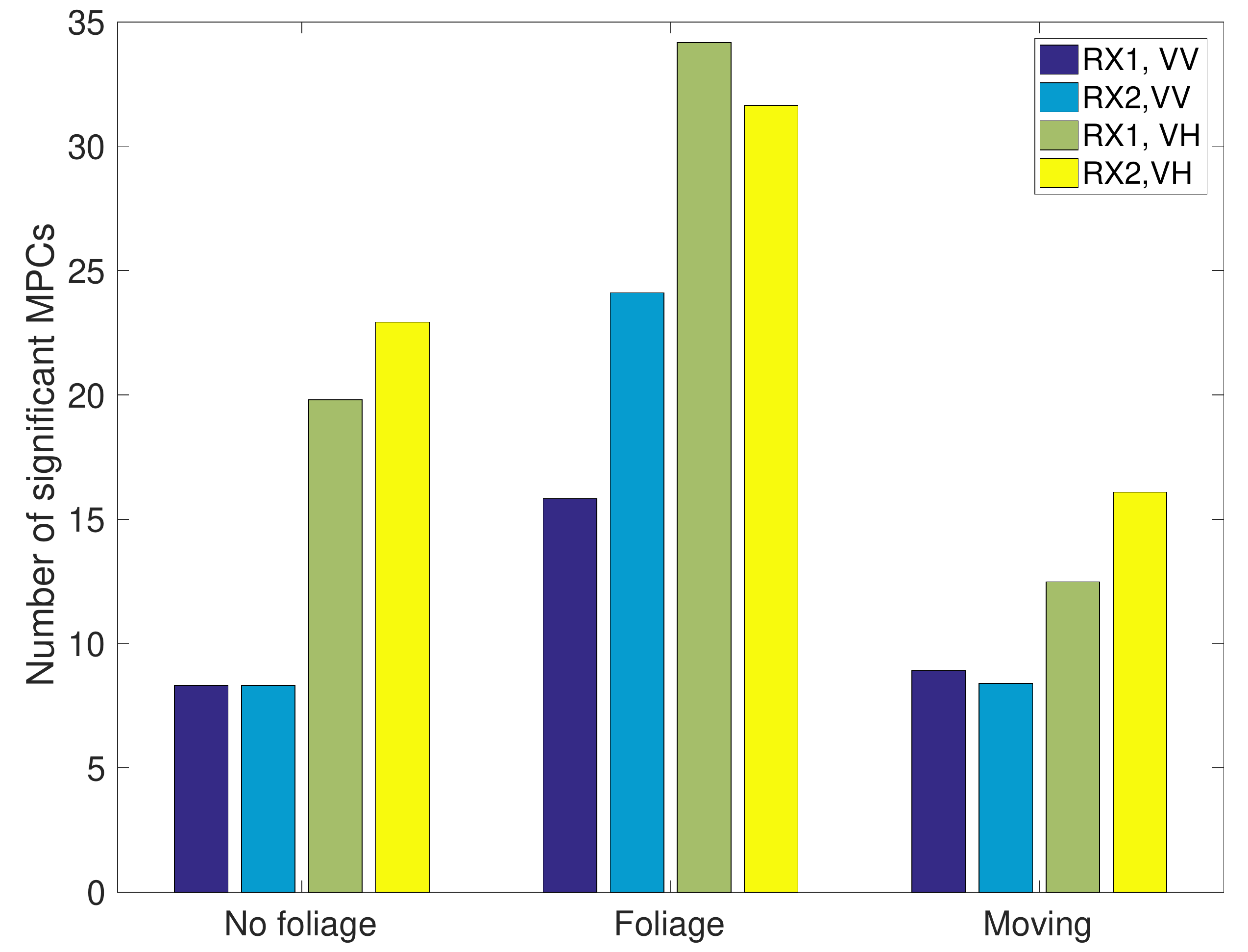}
	\caption{Average number of significant MPCs over multiple channel scans for (i) UAV hovering without foliage, (ii) with foliage obstruction, and (iii) UAV moving in a circle. The average number of significant MPCs are obtained by averaging over UAV heights at respective horizontal distances for receiver positions RX1 and RX2 with VV and VH antenna orientation pairs.}\label{Fig:Num_MPCs_mean}
\end{figure}

Fig.~\ref{Fig:PL_mean}(a) and Fig.~\ref{Fig:PL_mean}(b) show the path loss for the VV antenna orientation pair at RX1 and RX2, respectively. At RX1 with VV antenna orientation, for UAV hovering without foliage shown in red, we observe higher path loss at UAV height of $10$~m for $x=30$~m than for $x=15$~m due to larger link distance. However, as the UAV height increases, the path loss at $x=15$~m increases faster than that at $x=30$~m. As a result, the path loss at UAV height of $30$~m becomes higher for the UAV at $x=15$~m. This increase in the path loss is due to the smaller antenna gain at higher elevation angles as discussed in Section~\ref{Section:Antenna_model}. Similar observations were made at RX2 for this scenario.

The path loss for UAV moving scenario is shown in black, where we observe higher path loss compared to UAV hovering without foliage scenario. This can be explained by considering the antenna gain in the azimuth plane shown in Fig.~\ref{Fig:Antenna_pattern}. Due to motion of the UAV in a circular path, the RX antenna gains will change continuously in the azimuth and elevation planes. In the azimuth plane, the gain of antenna is smaller at $0^\circ$ and $180^\circ$. This UAV circular motion results in overall reduction in the path loss compared to the UAV hovering scenario, where the antennas are facing at boresight~($90^\circ$) all the time. The antenna gain in the elevation plane has slightly higher values at certain elevation angles compared to others as can be observed in Fig.~\ref{Fig:Elevation_gain_sine}. However, considering large number of slices of the antenna gain pattern in the elevation plane forming the overall three dimensional donut pattern, the antenna gain variations in the elevation plane will be averaged out during the UAV motion. This phenomenon also leads to closely spaced path loss curves for UAV moving scenario at $x=15$~m and $x=30$~m for three UAV heights. Similar observation can be are made at RX2 with results illustrated in Fig.~\ref{Fig:PL_mean}(b).

The path loss for VH antenna orientation pair is shown in Fig.~\ref{Fig:PL_mean}(c) and Fig.~\ref{Fig:PL_mean}(d) for RX1 and RX2, respectively. The path loss is higher than the VV antenna orientation at RX1 and RX2 due to polarization loss factor as discussed in Section~\ref{Section:Polarization}. However, the effect of the antenna gain in the elevation plane is negligible compared to VV antenna orientation due to large polarization mismatch loss. An interesting observation is that the polarization mismatch has smaller effect for UAV moving scenario than for the UAV hovering scenario without foliage. This is due to a larger number of cross polarized components arising during the UAV motion than when the UAV is hovering. For VH antenna orientation, the boresight of the antenna is facing the ground~(direction of azimuth emission) shown in Fig.~\ref{Fig:Scenario_AG}. This results in specular and diffuse reflections from the ground, that results in cross polarized components reaching the RXs. The motion of the UAV results in more cross polarized components~(assuming the ground surface is not uniform) compared to when the UAV is hovering. This results in higher received power. This phenomenon is more evident at $x=15$~m compared to $x=30$~m. The weaker cross polarized components generated at $x=15$~m are better received than at $x=30$~m.  

The path loss for UAV hovering without foliage scenario shows the highest change experienced due to antenna orientation misalignment~VH compared to when aligned~VV. On the other hand, for the foliage obstructed UAV hovering scenario, the path loss approximately remains constant. This is mainly due to the obstruction of the dominant LOS path, where, the received power of the LOS path is significantly larger~(approximately $20$~dB) than the other MPCs. The path loss for the foliage obstructed scenario does not show any significant effect of antenna orientation or link distance at RX1 and RX2. The path loss only shows a change at RX1 for UAV height of $10$~m at $x=15$~m. This is mainly due to the smaller obstruction~(thin tree trunk) experienced by the RX placed on the ground from a low altitude of the UAV. However, for RX2, and larger UAV heights, the tree crown obstructs the path. Moreover, at horizontal distance of $x=30$~m, the visibility of the RXs from the UAV become better, however, the path loss is larger due to larger link distance. Therefore, the path loss remains approximately constant.     

Overall, in addition to link distance $d$ between the GS and the UAV, the path loss is dependent on the elevation angle between the TX and the RX~(\ref{Eq:Received_power}), and antenna orientation--see ~Section~\ref{Section:Polarization}. The larger the elevation angle and orientation mismatch, the greater the path loss. This effect is more prominent when the antennas have the same orientation. Moreover, the foliage introduces additional path loss due to partial obstruction of the direct LOS. Moreover, the path loss at RX1 is higher than at RX2 for the UAV hovering without foliage and UAV moving scenarios for both VV and VH antenna orientation pairs. This is mainly due to higher ground absorption and at RX1 compared to RX2. This absorption is particularly higher for UWB radio signals due to better ground penetration properties. On the other hand, RX2 has better ground clearance and angle of reception than RX1.

\subsection{Multipath Channel Analysis: Number of Significant MPCs} \label{Section:Significant_MPCs}
The number of significant MPCs is obtained by selecting only the MPCs that are above the threshold of $20\%$ of the maximum amplitude for a given CIR. Time averaging them over the CIRs provides an average number of significant MPCs. The potential scatterers near the RXs that provided the MPCs are tripod, laptop, two humans and nearby sitting desks. The plot of the average number of significant MPCs is shown in Fig.~\ref{Fig:Num_MPCs_mean}. It can be observed that we have a larger number of significant MPCs for VH antenna orientation than for the VV antenna orientation pair. This is mainly due to low powered cross polarized components arising from specular and diffuse reflections. On the other hand, for the VV antenna orientation pair, the strength of the LOS component~(due to co-polarized electric fields) is far greater compared to the MPCs. Therefore, these low powered cross polarized components cannot be detected at the RX and would be below the noise floor~(due to limited dynamic range of the measurements). Similarly, we observe the highest number of significant MPCs for the foliage obstructed scenario, due to additional MPCs from the tree foliage. However, we have the smallest number of significant MPCs for the UAV moving scenario. This may simply be due to the sparsity of the channel along with the general antenna misalignment during flight. Overall, we observe larger number of MPCs for RX2 compared to RX1 mainly due to better ground clearance.  

\begin{table*}[!h]
\centering
\begin{tabular}{|p{0.2cm}|p{0.2cm}|p{0.2cm}|p{0.2cm}|p{0.2cm}|p{0.2cm}|p{0.2cm}|p{0.2cm}|p{0.2cm}|}
\hline
		\multicolumn{1}{|c|}{}&\multicolumn{2}{|c|}{\textbf{RX1~(VV)}}&\multicolumn{2}{|c|}{\textbf{RX2~(VV)}}&\multicolumn{2}{|c|}{\textbf{RX1~(VH)}}&\multicolumn{2}{|c|}{\textbf{RX2~(VH)}} \\
		
			\hline
             \multicolumn{1}{|c|}{\textbf{Param.}}&\multicolumn{1}{|c|}{\textbf{x = 15m}}&\multicolumn{1}{|c|}{\textbf{x = 30m}}&\multicolumn{1}{|c|}{\textbf{x = 15m}}&\multicolumn{1}{|c|}{\textbf{x = 30m}}&\multicolumn{1}{|c|}{\textbf{x = 15m}}&\multicolumn{1}{|c|}{\textbf{x = 30m}}&\multicolumn{1}{|c|}{\textbf{x = 15m}}&\multicolumn{1}{|c|}{\textbf{x = 30m}}\\
\hline
\multicolumn{1}{|c|}{$N_{\rm C}$}& \multicolumn{1}{|c|}{3.33}& \multicolumn{1}{|c|}{4}& \multicolumn{1}{|c|}{2.66}& \multicolumn{1}{|c|}{2} & \multicolumn{1}{|c|}{1.66}& \multicolumn{1}{|c|}{2.66}& \multicolumn{1}{|c|}{1.66}& \multicolumn{1}{|c|}{1.33}\\
\hline
\multicolumn{1}{|c|}{$\chi~(\frac{1}{\rm ns})$}& \multicolumn{1}{|c|}{.033}& \multicolumn{1}{|c|}{.04}& \multicolumn{1}{|c|}{.027}& \multicolumn{1}{|c|}{.02} & \multicolumn{1}{|c|}{.017}& \multicolumn{1}{|c|}{.027}& \multicolumn{1}{|c|}{.017}& \multicolumn{1}{|c|}{.013}\\
\hline
\multicolumn{1}{|c|}{$\eta$}& \multicolumn{1}{|c|}{.23}& \multicolumn{1}{|c|}{.186}& \multicolumn{1}{|c|}{.24}& \multicolumn{1}{|c|}{.16}& \multicolumn{1}{|c|}{.215}& \multicolumn{1}{|c|}{.16}& \multicolumn{1}{|c|}{.177}& \multicolumn{1}{|c|}{.171}\\
\hline
\multicolumn{1}{|c|}{$\varsigma~(\frac{1}{ns})$}& \multicolumn{1}{|c|}{.1}& \multicolumn{1}{|c|}{.06}& \multicolumn{1}{|c|}{.11}& \multicolumn{1}{|c|}{.06}& \multicolumn{1}{|c|}{.25}& \multicolumn{1}{|c|}{.15}& \multicolumn{1}{|c|}{.26}& \multicolumn{1}{|c|}{.2}\\
\hline
\multicolumn{1}{|c|}{$\gamma$}& \multicolumn{1}{|c|}{8.7}& \multicolumn{1}{|c|}{8.66}& \multicolumn{1}{|c|}{5.5}& \multicolumn{1}{|c|}{4.3}& \multicolumn{1}{|c|}{2.7}& \multicolumn{1}{|c|}{5.92}& \multicolumn{1}{|c|}{2.8}& \multicolumn{1}{|c|}{1.88}\\
\hline
		\end{tabular}
\caption{UWB UAV channel model parameters averaged over UAV heights for open area while the UAV is hovering without obstruction.} \label{Table:Parameters_open}
\end{table*}

\begin{table*}[!h]
\centering
\begin{tabular}{|p{0.2cm}|p{0.2cm}|p{0.2cm}|p{0.2cm}|p{0.2cm}|p{0.2cm}|p{0.2cm}|p{0.2cm}|p{0.2cm}|}
\hline
			\multicolumn{1}{|c|}{}&\multicolumn{2}{|c|}{\textbf{RX1~(VV)}}&\multicolumn{2}{|c|}{\textbf{RX2~(VV)}}&\multicolumn{2}{|c|}{\textbf{RX1~(VH)}}&\multicolumn{2}{|c|}{\textbf{RX2~(VH)}} \\
			
			\hline
             \multicolumn{1}{|c|}{\textbf{Param.}}&\multicolumn{1}{|c|}{\textbf{x = 15m}}&\multicolumn{1}{|c|}{\textbf{x = 30m}}&\multicolumn{1}{|c|}{\textbf{x = 15m}}&\multicolumn{1}{|c|}{\textbf{x = 30m}}&\multicolumn{1}{|c|}{\textbf{x = 15m}}&\multicolumn{1}{|c|}{\textbf{x = 30m}}&\multicolumn{1}{|c|}{\textbf{x = 15m}}&\multicolumn{1}{|c|}{\textbf{x = 30m}}\\
\hline
\multicolumn{1}{|c|}{$N_{\rm C}$}& \multicolumn{1}{|c|}{2}& \multicolumn{1}{|c|}{2}& \multicolumn{1}{|c|}{2}& \multicolumn{1}{|c|}{1.66}& \multicolumn{1}{|c|}{2}& \multicolumn{1}{|c|}{1.33}& \multicolumn{1}{|c|}{1.66}& \multicolumn{1}{|c|}{1.33} \\
\hline
\multicolumn{1}{|c|}{$\chi~(\frac{1}{\rm ns})$}& \multicolumn{1}{|c|}{.02}& \multicolumn{1}{|c|}{.02}& \multicolumn{1}{|c|}{.02}& \multicolumn{1}{|c|}{.017}& \multicolumn{1}{|c|}{.02}& \multicolumn{1}{|c|}{.013}& \multicolumn{1}{|c|}{.017}& \multicolumn{1}{|c|}{.013} \\
\hline
\multicolumn{1}{|c|}{$\eta$}& \multicolumn{1}{|c|}{.212}& \multicolumn{1}{|c|}{.21}& \multicolumn{1}{|c|}{.24}& \multicolumn{1}{|c|}{.23}& \multicolumn{1}{|c|}{.214}& \multicolumn{1}{|c|}{.16}& \multicolumn{1}{|c|}{.198}& \multicolumn{1}{|c|}{.2} \\
\hline
\multicolumn{1}{|c|}{$\varsigma~(\frac{1}{ns})$}& \multicolumn{1}{|c|}{.14}& \multicolumn{1}{|c|}{.175}& \multicolumn{1}{|c|}{.27}& \multicolumn{1}{|c|}{.21}& \multicolumn{1}{|c|}{.34}& \multicolumn{1}{|c|}{.34}& \multicolumn{1}{|c|}{.3}& \multicolumn{1}{|c|}{.34} \\
\hline
\multicolumn{1}{|c|}{$\gamma$}& \multicolumn{1}{|c|}{1.3}& \multicolumn{1}{|c|}{1.11}& \multicolumn{1}{|c|}{.985}& \multicolumn{1}{|c|}{1.34}& \multicolumn{1}{|c|}{.77}& \multicolumn{1}{|c|}{.811}& \multicolumn{1}{|c|}{1.4}& \multicolumn{1}{|c|}{.74} \\
\hline
		\end{tabular}
\caption{UWB UAV channel model parameters averaged over UAV heights for open area obstructed by foliage while the UAV is hovering.}\label{Table:Parameters_foliage}
\end{table*}

\begin{table*}[!h]
\centering
\begin{tabular}{|p{0.2cm}|p{0.2cm}|p{0.2cm}|p{0.2cm}|p{0.2cm}|p{0.2cm}|p{0.2cm}|p{0.2cm}|p{0.2cm}|}
\hline
			\multicolumn{1}{|c|}{}&\multicolumn{2}{|c|}{\textbf{RX1~(VV)}}&\multicolumn{2}{|c|}{\textbf{RX2~(VV)}}&\multicolumn{2}{|c|}{\textbf{RX1~(VH)}}&\multicolumn{2}{|c|}{\textbf{RX2~(VH)}} \\
			\hline
             \multicolumn{1}{|c|}{\textbf{Param.}}&\multicolumn{1}{|c|}{\textbf{x = 15m}}&\multicolumn{1}{|c|}{\textbf{x = 30m}}&\multicolumn{1}{|c|}{\textbf{x = 15m}}&\multicolumn{1}{|c|}{\textbf{x = 30m}}&\multicolumn{1}{|c|}{\textbf{x = 15m}}&\multicolumn{1}{|c|}{\textbf{x = 30m}}&\multicolumn{1}{|c|}{\textbf{x = 15m}}&\multicolumn{1}{|c|}{\textbf{x = 30m}}\\
\hline
\multicolumn{1}{|c|}{$N_{\rm C}$}& \multicolumn{1}{|c|}{2}& \multicolumn{1}{|c|}{1.66}& \multicolumn{1}{|c|}{1.66}& \multicolumn{1}{|c|}{1.33}& \multicolumn{1}{|c|}{2}& \multicolumn{1}{|c|}{1}& \multicolumn{1}{|c|}{1.66}& \multicolumn{1}{|c|}{1}\\
\hline
\multicolumn{1}{|c|}{$\chi~(\frac{1}{\rm ns})$}& \multicolumn{1}{|c|}{.02}& \multicolumn{1}{|c|}{.017}& \multicolumn{1}{|c|}{.017}& \multicolumn{1}{|c|}{.013}& \multicolumn{1}{|c|}{.02}& \multicolumn{1}{|c|}{.01}& \multicolumn{1}{|c|}{.017}& \multicolumn{1}{|c|}{.01}\\
\hline
\multicolumn{1}{|c|}{$\eta$}& \multicolumn{1}{|c|}{.14}& \multicolumn{1}{|c|}{.143}& \multicolumn{1}{|c|}{.2}& \multicolumn{1}{|c|}{.18}& \multicolumn{1}{|c|}{.15}& \multicolumn{1}{|c|}{.12}& \multicolumn{1}{|c|}{.205}& \multicolumn{1}{|c|}{.171}\\
\hline
\multicolumn{1}{|c|}{$\varsigma~(\frac{1}{ns})$}& \multicolumn{1}{|c|}{.1}& \multicolumn{1}{|c|}{.082}& \multicolumn{1}{|c|}{.084}& \multicolumn{1}{|c|}{.084}& \multicolumn{1}{|c|}{.14}& \multicolumn{1}{|c|}{.11}& \multicolumn{1}{|c|}{.16}& \multicolumn{1}{|c|}{.16}\\
\hline
\multicolumn{1}{|c|}{$\gamma$}& \multicolumn{1}{|c|}{1.87}& \multicolumn{1}{|c|}{1.87}& \multicolumn{1}{|c|}{3.6}& \multicolumn{1}{|c|}{5.2}& \multicolumn{1}{|c|}{1.76}& \multicolumn{1}{|c|}{2}& \multicolumn{1}{|c|}{2.04}& \multicolumn{1}{|c|}{1.31}\\
\hline
		\end{tabular}
\caption{UWB UAV channel model parameters averaged over UAV heights for open area while the UAV is moving.}\label{Table:Parameters_moving}
\end{table*}

\subsection{Multipath Channel Analysis: Channel Model Parameters}
The channel model parameters are obtained from empirical results and modeled mathematically in (\ref{Eq:Eq_CIR}), (\ref{Eq:Eq_Pwr}), (\ref{Eq:Eq_Arrival_cluster}) and (\ref{Eq:Eq_Arrival_MPC}), in Section~\ref{Section:CIR}. In this subsection, we discuss statistical propagation channel model parameters obtained from these equations. These parameters are provided in Table~\ref{Table:Parameters_open}, Table~\ref{Table:Parameters_foliage}, and Table~\ref{Table:Parameters_moving} for the three scenarios. It can be observed that the cluster arrival rate $\chi$ captured in (\ref{Eq:Eq_Arrival_cluster}) is the highest for the UAV hovering scenario without foliage, followed by foliage obstructed and UAV moving scenarios. Similarly, $\chi$ is higher for VV antenna orientation pair than for the VH antenna orientation pair for the three scenarios at both RX1 and RX2. This is mainly due to larger number of clusters observed during the excess delay duration, which will be explained in Section~\ref{Section:Power_Clusters}. 

The arrival rate of MPCs $\varsigma$ in (\ref{Eq:Eq_Arrival_MPC}) is the highest for foliage obstructed scenario compared to the other two scenarios. This is mainly due to multiple reflections from the tree foliage~(branches, leaves and trunk). Similarly, $\varsigma$ is higher for VH antenna orientation pair at both RX1 and RX2 than for the VV antenna orientation pair, as explained in Section~\ref{Section:Significant_MPCs}.

The cluster power decay constant $\eta$ shown in (\ref{Eq:Eq_Pwr}) is approximately similar for all three scenarios. The clusters in different scenarios have specific power rise and fall values during respective time bins. For example, for the UAV hovering without foliage scenario, we observe smaller decay of cluster power over small time bins than in the foliage obstructed scenario, where we have larger power decay over larger time bins. However, we observe higher $\eta$ for the VV antenna orientation pair than for the VH antenna orientation pair at both RX1 and RX2. This is mainly due to larger power variation across the clusters for the VV antenna orientation compared to the VH antenna orientation, where the clusters are nearly all low powered. Moreover, for the UAV moving scenario, we observe the smallest overall $\eta$, showing more consistent received power, (averaged over the channel scans) when the UAV is moving at a uniform speed in a circle at a constant link distance. 

Important observations can be made from the power decay constant of the MPCs, $\gamma$ obtained in (\ref{Eq:Eq_Pwr}), in three different scenarios. The value of $\gamma$ is the highest for the UAV hovering without foliage scenario. The clusters observed in this case are of short duration with sharp power decay~(see Section~\ref{Section:Power_Clusters}). This results in larger $\gamma$ value for the MPCs within the clusters. On the other hand, in case of foliage, we observe only few clusters. The power from the large duration foliage clusters decays slowly, resulting in overall smaller $\gamma$ . The value of $\gamma$ for the UAV moving scenario is in between the two previously mentioned scenarios. Moreover, $\gamma$ is larger for the VV antenna orientation pair than for the VH antenna orientation pair at both RX1 and RX2. This indicates that power of MPCs decays sharply for VV antenna orientation when compared to the VH antenna orientation, where the power from cross polarized components does not change significantly.

\subsection{Multipath Channel Analysis: Power Clusters} \label{Section:Power_Clusters}
A common phenomenon to observe during UWB propagation is the clustered reception of power~\cite{uwb_cluster}. This is mainly due to the fine time resolution of individual MPCs reflected from a scatterer, whose combined power can be viewed as a cluster during a given time bin. In our outdoor open area environment with limited obstacles and excess delay, we observe a small number of clusters in the PDP shown in Fig.~\ref{Fig:PDP}. The clusters are identified by visual inspection. The clusters are identified based on the distinct boundaries of power fall and rise. Whenever, we have a change of the slope of the power decay shown in Fig.~\ref{Fig:PDP}, with a threshold of at least $10$~dB, and the duration from the peak to the fall~(change of slope again) is at least $2$~ns, a cluster is said to be formed. The mean cluster count $N_{\rm C}$ captured in (\ref{Eq:Eq_CIR}), is provided in Table~\ref{Table:Parameters_open}, Table~\ref{Table:Parameters_foliage}, and Table~\ref{Table:Parameters_moving}. The UAV hovering without foliage scenario has the largest mean cluster count followed by the foliage obstructed scenario and UAV moving scenarios for different link propagation settings provided in Table~\ref{Table:Parameters_open}, Table~\ref{Table:Parameters_foliage}, and Table~\ref{Table:Parameters_moving}, respectively.  

In case of UAV hovering without obstruction scenario shown in Fig.~\ref{Fig:Scenario_AG}(a), we observe essentially independent reflections from small scatterers near the RX, yielding several distinct clusters. However, in case of UAV hovering with link obstructed by tree foliage scenario, shown in Fig.~\ref{Fig:Scenario_AG}(b), mainly $2$ clusters are observed: One due to OLOS and the second from the tree body and other nearby scatterers. In addition, in case of foliage, the second cluster time bin is large compared to the other scenarios due to multiple reflections from foliage. On the other hand, in case of UAV moving scenario in open area shown in Fig.~\ref{Fig:Scenario_AG}(c), we have a small number of clusters due to motion.

 We observe larger mean cluster count for the VV antenna orientation pair compared to the VH antenna orientation pair for all the propagation scenarios. This behavior is mainly due to polarization misalignment at the RX for VH antenna orientation. The received power in case of VH antenna orientation changes less rapidly compared to VV antenna orientation, leading to smaller cluster count. Similarly, we observe larger mean cluster count at RX1 than RX2. This is due to reflections from the tripod body. The tripod body provides additional reflections and at the same time may help guide the energy towards the RX on the ground shown in Fig.~\ref{Fig:Scenario_original}.

\section{Conclusions}\label{Section:Conclusions} 
In this work, we have conducted UWB AG propagation channel measurements in an open field using a small UAV in three propagation scenarios: UAV hovering without foliage, foliage obstruction, and UAV moving in a circle. Measurements were obtained at three UAV heights and two horizontal distances for two different antenna orientations at the UAV TX. We observed that the received power is highly dependent on the antenna gain of the LOS component in the elevation plane when the antennas are aligned~(same orientation). Moreover, the antenna gain for the LOS component can be approximated by a sine function of the elevation angle between the TX and the RX. Antenna orientation mismatch results in higher path loss and larger number of MPCs (as expected). The OLOS scenario due to foliage between the TX and the RX while the UAV is hovering introduces additional attenuation, and additional MPCs due to foliage (also as expected). Moreover, the motion of the UAV in the circular path provides better mitigation against antenna orientation mismatch compared to the UAV hovering scenario without foliage. A statistical channel model was obtained from the empirical results which show that larger number of MPC clusters were observed for the UAV hovering without foliage, than for the UAV moving scenario.

\section*{Acknowledgement}
This  work  has  been  supported  in  part  by  NASA  under  the  Federal Award ID number NNX17AJ94A and by National Science Foundation~(NSF) under the grant number CNS-1453678. Wahab Khawaja has also been supported via a Fulbright scholarship. We also thank Jianlin Chen from NCSU, for his help in the measurements. 

\bibliographystyle{IEEEtran}



\begin{biographywithpic}{Wahab Khawaja}{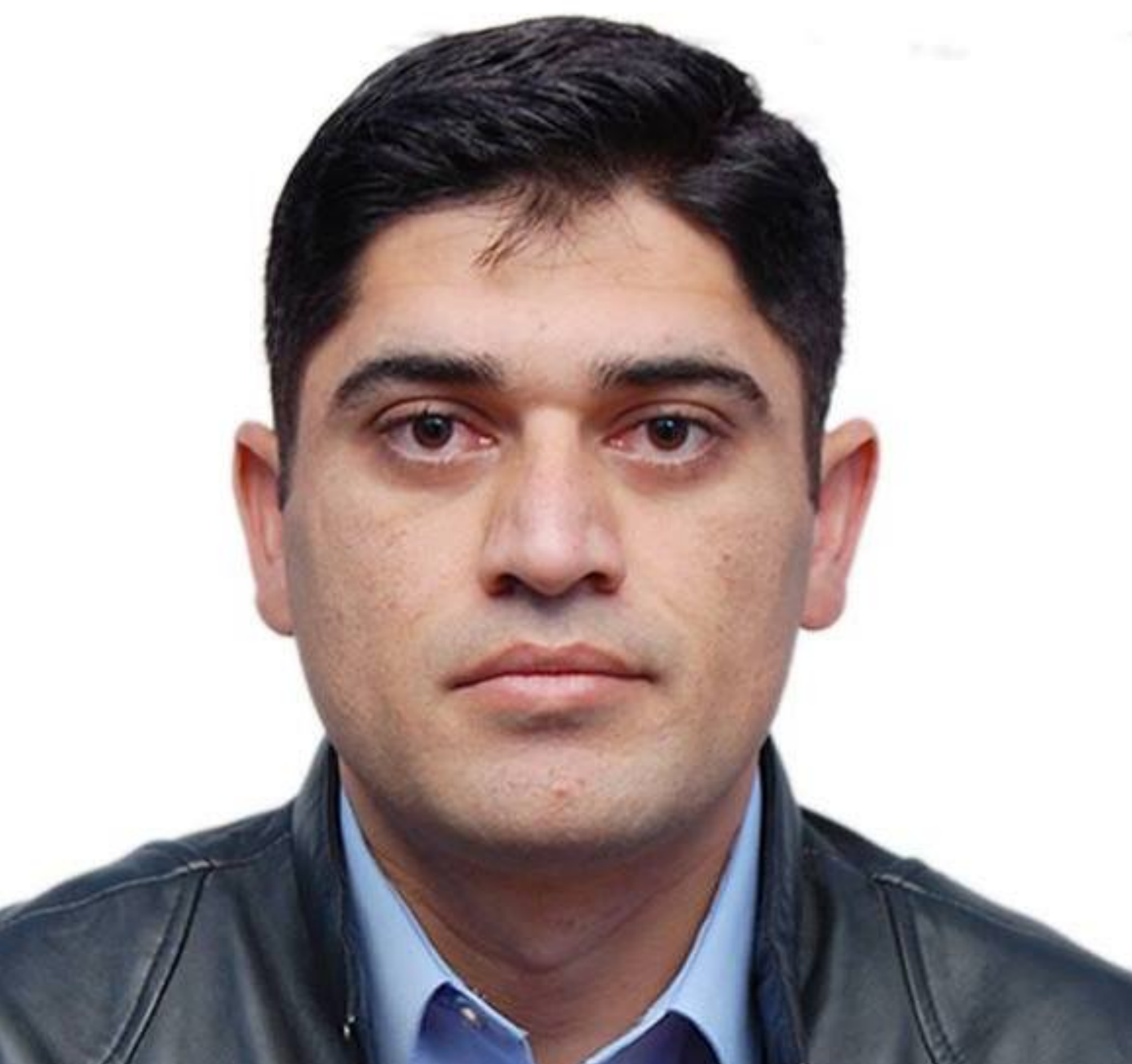}
received his BS and MS degrees both in electrical engineering from University of Engineering and Technology, Lahore, Pakistan in 2008 and 2012, respectively. He is currently pursuing his PhD degree at Electrical and Computer Engineering department of NCSU on Fulbright scholarship. His current research areas are UWB channel modeling, mmWave channel modeling using UAVs or static platforms in different environmental conditions. 
\end{biographywithpic}

\begin{biographywithpic}{Ozgur Ozdemir}{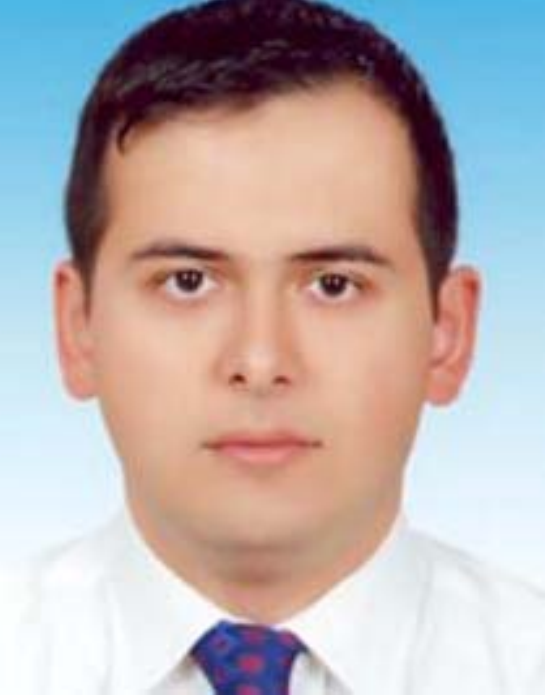}
received the BS degree in electrical and electronics engineering from Bogazici University, Istanbul, Turkey, in 1999 and the MS and PhD degrees in electrical engineering from University of Texas at Dallas, Richardson, TX, USA, in 2002 and 2007, respectively. Before joining to Department of Electrical and Computer Engineering at NCSU as a visiting research scholar, he has been an assistant professor at the Department of Electrical and Electronics Engineering, Fatih University, Turkey and a postdoctoral scholar at Qatar University, Doha, Qatar.His research interest include opportunistic approaches in wireless systems, experimental multiple-antenna systems, digital compensation of radio-frequency impairments, wireless multi-carrier communications, software defined radios.
\end{biographywithpic}

\begin{biographywithpic}{Fatih Erden}{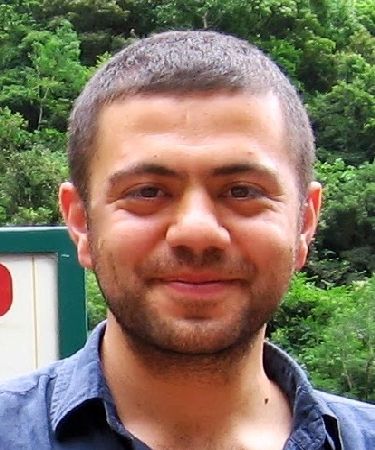}
received the B.S. and M.S. degrees from Bilkent University, Ankara, Turkey, in 2007 and 2009, respectively, and the Ph.D. degree from Hacettepe University, Ankara, Turkey, in 2015, all in electrical and electronics engineering. 
From 2015 to 2016, he was an Assistant Professor with the Dept. of Electrical and Electronics Eng., Atilim University, Ankara, Turkey. From 2016 to 2018, he was with the Signal Processing Group at Bilkent University, Ankara, Turkey as a postdoctoral researcher. He is now working as a research associate at the Dept. of Electrical and Computer Eng. at North Carolina State University. 
His research interests include signal and image processing, pattern recognition, time-series analysis, infrared sensors, sensor fusion, and EV-grid integration.

\begin{biographywithpic}{Ismail Guvenc}{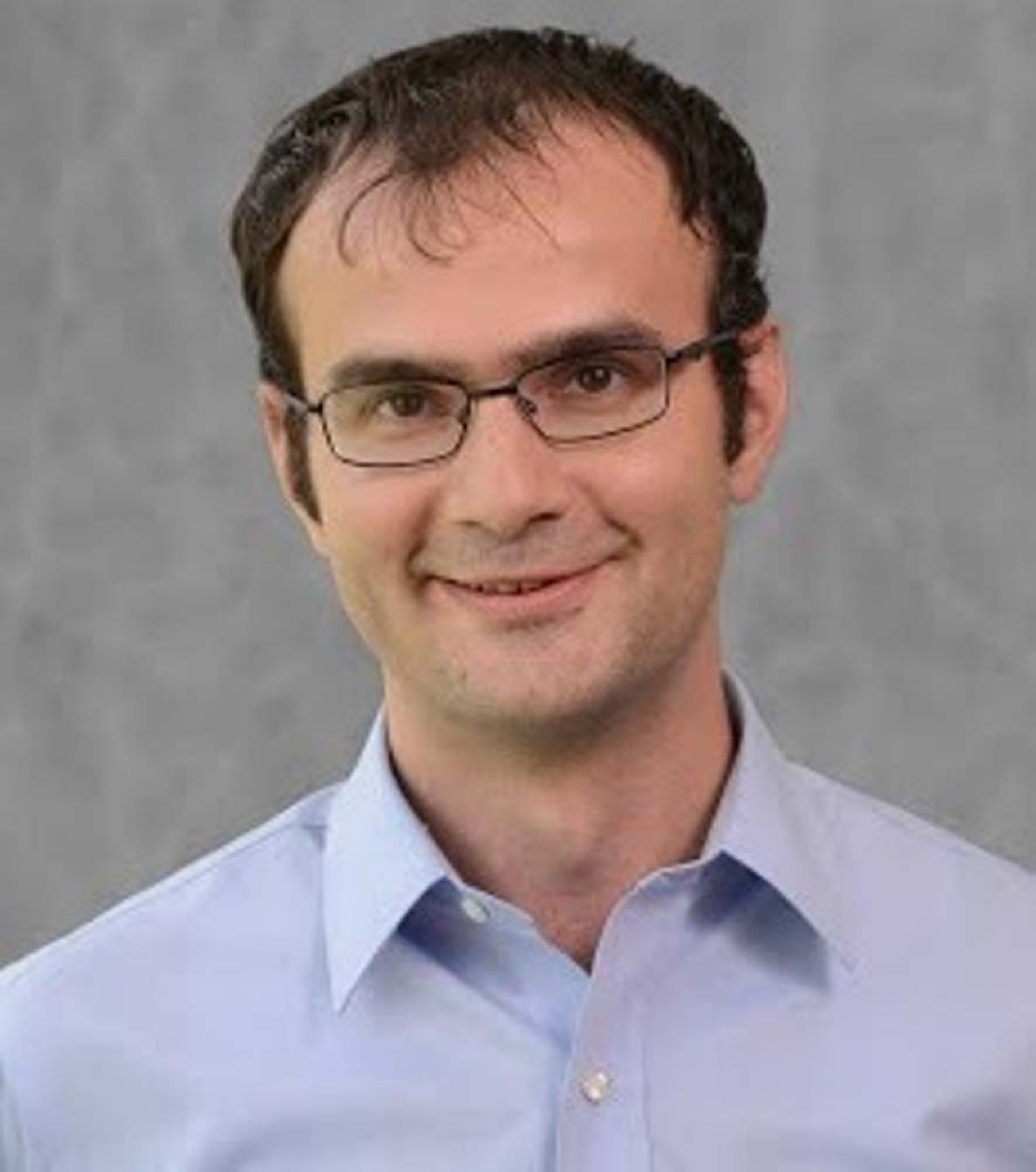}
has been an Associate Professor at North Carolina State University since August 2016. Before joining NCSU, he has worked as a research engineer at DOCOMO Innovations, Inc. in Palo Alto, CA (2006-2012) and as an assistant professor at Florida International University in Miami, FL (2012-2016). His recent research interests include 5G and mmWave wireless networks, UAV communications, and heterogeneous networks.. He is a recipient of the 2016 FIU College of Engineering Faculty Research Award, 2015 NSF CAREER Award, 2014 Ralph E. Powe Junior Faculty Award, and 2006 USF Outstanding Dissertation Award.
\end{biographywithpic}

\begin{biographywithpic}{David Matolak}{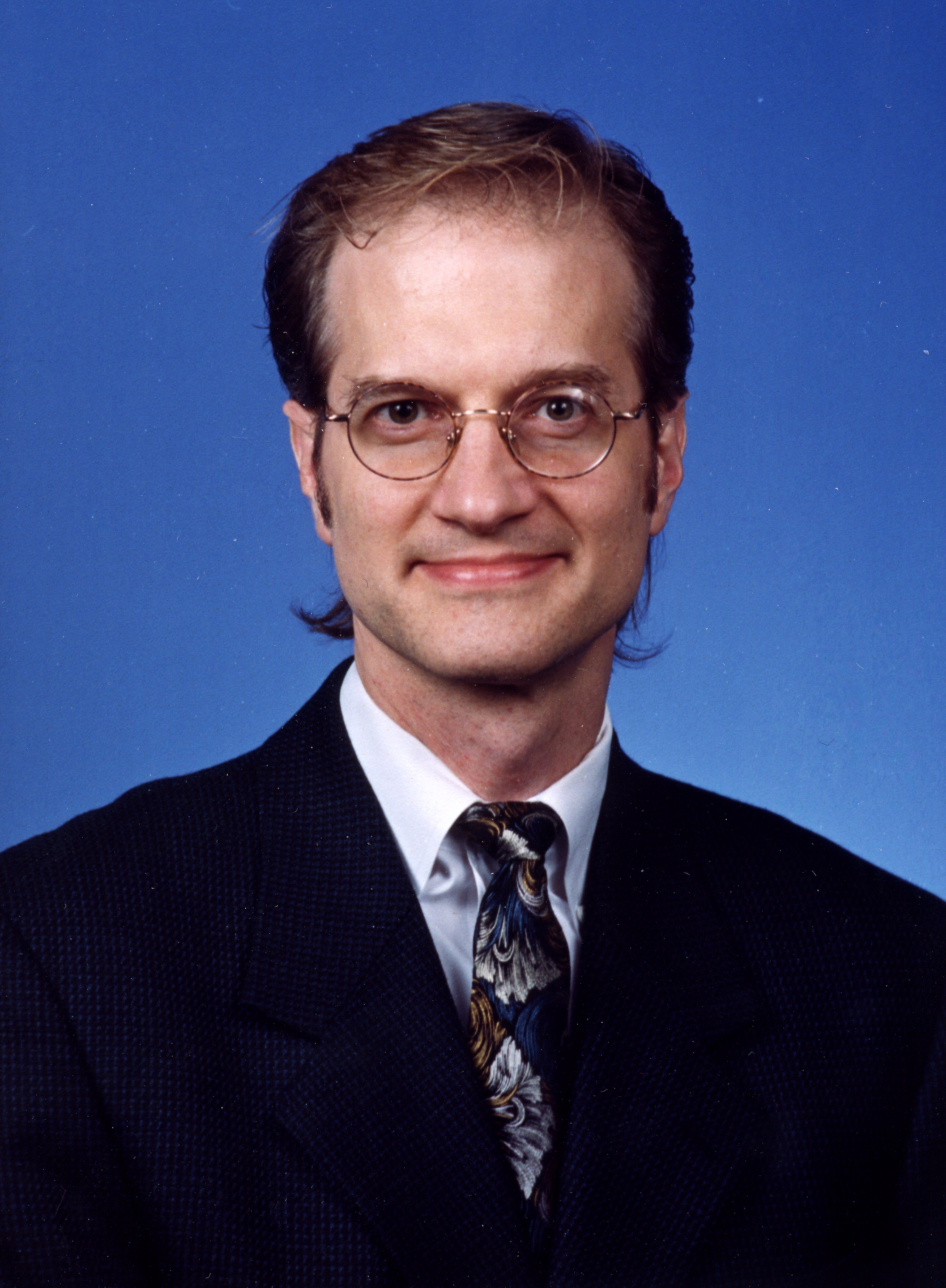}
received his M.S., in Electrical and Computer Engineering, from University of Massachusetts, Amherst, MA, in 1987 and received his Ph.D., in Electrical Engineering, from University of Virginia, Charlottesville, VA, in 1995. He is currently serving as a Professor in Electrical and Computer Eng. Dept., South Carolina State University. His research interests include
Wireless Channel Characterization, Ad Hoc Communication Networks, Modulation/Detection PHY/MAC, Random Process Modeling, and Electromagnetics Modeling. 
\end{biographywithpic}

\end{biographywithpic}

\end{document}